\documentclass[a4paper,fleqn,usenatbib,useAMS]{mnras}

\pdfoutput=1
\usepackage[T1]{fontenc}
\usepackage{aecompl}

\usepackage[pdftex]{graphicx}
\usepackage{amsmath}	
\usepackage{amssymb}	
\usepackage{threeparttable}
\usepackage{bm}		
\usepackage{pdflscape}	
\usepackage{ulem}

\newcommand{\solmass}{\rm M_{\odot}}

\providecommand{\adsurl}[1]{\href{#1}{ADS}}

\newcommand{\per}[2]{\rm {#1}^{-#2}}

\newcommand{\Lya}{$\rm{Ly} \alpha$ }

\title[{\sf SEURAT}: line transfer in SPH]
{{\sf SEURAT}: SPH scheme extended with ultraviolet line radiative transfer}
\author[M. Abe et al. ]{Makito Abe$^{1}$\thanks{E-mail: mabe@ccs.tsukuba.ac.jp (MA)}, Hiroyuki Suzuki$^{1}$, Kenji Hasegawa$^{2}$, Benoit Semelin$^{3}$,\newauthor Hidenobu Yajima$^{4,}$ $^{5}$ and Masayuki Umemura$^{1}$ \\
$^{1}$Center for Computational Sciences, University of Tsukuba, Ten-nodai, 1-1-1 Tsukuba, Ibaraki 305-8577, Japan\\
$^{2}$Graduate School of Science, Nagoya University, Furo-cho, Chikusa-ku, Nagoya, Aichi 464-8602, Japan\\
$^{3}$LERMA, Observatoire de Paris, Sorbonne Universit\'e, PSL research university, CNRS, F-75014, Paris, France\\
$^{4}$Frontier Research Institute for Interdisciplinary Sciences, Tohoku University, Sendai 980-8578, Japan\\
$^{5}$Astronomical Institute, Tohoku University, Sendai 980-8578, Japan}

\begin{document}
\label{firstpage}
\pagerange{\pageref{firstpage}--\pageref{lastpage}}
\maketitle


\begin{abstract}
We present a novel Lyman alpha (Ly$\alpha$) radiative transfer code, {\small\sf SEURAT}, 
where line scatterings are solved adaptively with the resolution of the smoothed particle hydrodynamics (SPH). 
The radiative transfer method implemented in {\small\sf SEURAT} is based on a Monte Carlo algorithm in which the scattering and absorption by dust are also incorporated.  
We perform standard test calculations to verify the validity of the code; 
(i) emergent spectra from a static uniform sphere,  (ii) emergent spectra from an expanding uniform sphere, 
and (iii) escape fraction from a dusty slab. 
Thereby we demonstrate that our code solves the \Lya radiative transfer with sufficient accuracy. 
We emphasise that {\small\sf SEURAT} can treat the transfer of \Lya photons even in highly complex systems 
that have significantly inhomogeneous density fields. 
The high adaptivity of {\small\sf SEURAT} is desirable to solve the propagation of \Lya photons in the interstellar medium 
of young star-forming galaxies like \Lya emitters (LAEs). 
Thus, {\small\sf SEURAT} provides a powerful tool to model the emergent spectra of \Lya emission, which can be
compared to the observations of LAEs. 
\end{abstract}



\section{Introduction}
The hydrogen \Lya emission is a significant probe to explore high-$z$ young galaxies, 
as pioneeringly suggested by \citet{Partridge&Peebles67}. 
So far, a number of distant galaxies have been detected via intensive observations of the \Lya line, 
the so-called \Lya emitters \citep[LAEs:][]{Iye+06,Ouchi+08,Ouchi+10,Vanzella+11,Ono+12,Shibuya+12,Finkelstein+13,Konno+14,Zitrin+15}. 
Recently, the most distant LAE has been detected at $z=8.68$ by \citet{Zitrin+15}. 
It is difficult to reveal the detailed structure and physical properties of distant galaxies even with sate-of-the-art observational facilities. 
However, the \Lya line can in principle provide copious information about the internal structure of those galaxies, because the emergent spectra of \Lya line depend on velocity fields and ionization structure in the galaxies \citep{Dijkstra+06, Dijkstra+06b, Verhamme+06, ART2, Yajima+15} and the photon escape fraction is sensitive to the metallicity of the gas \citep{Atek+08,Verhamme+08,Yajima+14}. 

In addition, the neutral hydrogen in the intergalactic medium (IGM) erodes the \Lya emission from high-$z$ galaxies. 
Consequently, the cosmic reionization history is constrained by investigating the redshift evolution of the luminosity function (LF) of LAEs \citep[e.g.,][]{Kashikawa+06,Ouchi+10}. 
Moreover, LAEs can be the main ionizing sources responsible for the cosmic reionization \citep{Yajima+09, Yajima+14}. 
It also should be noted that the shape of  \Lya line emission can constrain the neutral fraction of the IGM 
by considering the distortion due to the IGM transmission \citep{Santos04,Dijkstra+07}. 
Therefore, understanding the \Lya emission from high-redshift galaxies 
is crucial for revealing not only galaxy formation but also cosmic reionization. 

\Lya photons are emitted by the 2$P$-1$S$ transitions as a result of the excitation of H{\sc i} atoms or
the recombination in H{\sc ii} regions \citep{Faucher-Giguere+10, Yajima+12}. 
Due to the large cross section of  \Lya radiation to neutral hydrogen, 
the interstellar medium (ISM) of LAEs can be readily optically thick to \Lya photons. 
In general, the transfer in a spectral line arising from the spontaneous bound-bound transition can be treated 
as resonant scattering, if the transition timescale is much shorter than the other physical timescales
(e.g. collisional deexcitation) and also the decaying to some other state is negligible. 
Since the \Lya transition with a large Einstein's A coefficient satisfies these conditions in the ISM of LAEs, 
we can treat the absorption and subsequent re-emission of a \Lya photon as a resonant scattering process. 
When scatterings are included, the radiative transfer equation takes the form of 
an integro-differential equation that should be solved through an iterative procedure. 
Moreover, 
we should pay special attention to the partial frequency redistribution 
during line scatterings, which is essential in Ly$\alpha$ radiative transfer. 
Therefore, in order to obtain an exact solution of the Ly$\alpha$ radiation transfer equation, 
we have to solve iteratively the frequency-dependent radiative transfer equation. 
Needless to say, it is generally time consuming to integrate such an equation directly. 
Under the diffusion approximation, analytical solutions for the emergent spectrum 
were derived for systems with simple geometry composed of pure hydrogen, such as a static 
and uniform slab or a spherical cloud \citep{Harrington73,Neufeld90, Dijkstra+06}. 
On the other hand, for inhomogeneous or moving gas clouds, \Lya properties have to be calculated by numerical simulations. 

A Monte Carlo approach, in which various phenomena are stochastically treated, is often 
employed to manage the complexity of the partial frequency redistribution in \Lya line transfer \citep[e.g.,][]{Zheng&Miralda02,Verhamme+06,LICORICE,Baek+09,Laursen+09,ART2,Smith+14,Yajima+14,Yajima&Li14}. 
Some previous works investigated the \Lya radiation properties of galaxies by using simple models \citep[e.g.,][]{Dijkstra+06,Dijkstra+11,Verhamme+06,Verhamme+08, Gronke+15}. 
For example, \citet{Verhamme+08} solved the \Lya radiation transfer in spherical expanding shells
and reproduced  the \Lya properties of observed galaxies with tuned H{\sc i} column densities and expansion velocities \citep[see also][]{Dijkstra+11}. 
However, the ISM can have a complicated structure due to stellar feedback, instability of the galactic disk, and interaction with other galaxies. 
Recent observations with high-angular resolution have revealed 
the inhomogeneous and clumpy ISM of high-$z$ galaxies \citep[e.g.,][]{Genzel+11}. 
Therefore, a solver which allows us to treat \Lya radiation transfer in more complex ISM structures is required. 


Recent cosmological simulations have been able to model galaxies and resolve their internal ISM structure \citep[e.g.,][]{Wise+12, Hasegawa&Semelin13,Hopkins+14,Vogelsberger+14,Schaye+15}. 
Most of simulations have been conducted using the smoothed particle hydrodynamics (SPH), 
which is a Lagrangian numerical scheme \citep[e.g., see a review by][]{Springel10}. 
An important advantage of SPH is that the spatial resolution automatically augments with increasing local density. 
Therefore, the SPH method can resolve adaptively the ISM and star-forming regions 
which produce \Lya photons. 
So far, Monte Carlo simulations of \Lya radiative transfer have been combined with SPH by mesh-based schemes,
where the SPH densities are assigned on the meshes before solving the \Lya radiative transfer \citep[e.g., ][]{LICORICE,ART2}. 
\citet{Laursen+09}, combining cosmological SPH simulations with post-processing \Lya transfer on the meshes, 
investigated the \Lya properties of high-$z$ galaxies and showed that the \Lya flux changes by a factor of $3-6$ 
depending on the viewing angles due to the complex ISM structure \citep[see also,][]{Yajima+12}. 
\citet{Yajima+15} calculated the \Lya properties of high-$z$ progenitors of a local Milky Way-like galaxy, 
and showed that the filamentary accreting gas produces \Lya photons efficiently via the excitation cooling process at $z > 6$.
However, the mesh-based schemes of  \Lya radiative transfer inevitably smooth out highly-resolved structures in SPH 
simulations, and might lead to a lack of accuracy in the \Lya photon transfer. 
In this paper, we develop a novel Monte Carlo scheme for \Lya radiative transfer adapted for SPH simulations, 
{\small\sf SEURAT} (SPH scheme Extended with Ultraviolet line RAdiative Transfer). 
Some basic parts of {\small\sf SEURAT} are the extension of the mesh-based \Lya radiation transfer code {\small\sf LICORICE} \citep{LICORICE}. 
In {\small\sf SEURAT}, SPH particles themselves are directly used to solve the radiation transfer 
unlike in the previous mesh-based codes. 
Hence, the transfer of \Lya photons can be pursued without reducing the resolution of SPH simulations. 


This paper is organized as follows. 
In Section 2, we describe the basic physics of \Lya radiation transfer and the algorithm of {\small\sf SEURAT}. 
The results of some standard test calculations are presented in section 3. 
In section 4, we demonstrate the adaptivity of  {\small\sf SEURAT} 
for highly inhomogeneous media. 
Also, we compare the \Lya transfer with {\small\sf SEURAT} to that with {\small\sf LICORICE} for
a model galaxy obtained by a cosmological radiation SPH simulation.
Section 5 is devoted to the conclusions.

\section{The code}

We firstly describe the flow of our Monte Carlo radiative transfer calculations. 
Once a photon packet (a monochromatic group of photons) is sent from a radiation source, the packet flies in a straight line along a stochastically determined direction until an interaction by an atom occurs (scattering or absorption). 
The probability distribution function of the path length that a packet can travel without an interaction is $\exp(-\tau)$, where $\tau$ is the optical depth of the path.   
In order to treat radiative interaction events stochastically, we randomly chose an optical depth $\tau_{\rm target}$, where an interaction takes place, as $\tau_{\rm target} = -\ln \xi$ with $\xi$ being a uniform random number between 0 and 1. 
We then integrate the optical depth along the light ray and let the photon packet propagate until $\tau = \tau_{\rm target}$ as described later in \S \ref{sec:RT}. 
When the packet reaches $\tau = \tau_{\rm target}$, we stochastically determine what happens there. 
Only scatterings occur in the pure hydrogen gas, while both absorption and scattering take place in dusty gas. 
If a scattering occurs at $\tau = \tau_{\rm target}$, 
we settle the resultant direction and frequency shift of the scattered photon packet following the method described in \S \ref{sec:physics}, and chose a new target optical depth. 
On the other hand, if the packet is absorbed, we completely eliminate the packet, 
or reduce the photon flux by a factor of attenuation. 
We repeat this flow until the photon packet escapes or vanishes from the system (\S \ref{sec:RT} and \ref{sec:dust}). 

\subsection{Treatment of frequency shifts during the Ly$\alpha$ scattering process}\label{sec:physics}
The broadening effect due to the quantum uncertainty results in the Lorentzian line profile $\phi(\nu)$ described as 
\begin{equation}
	\label{eq:Lorentzian}
	\phi(\nu) = \frac{\Delta \nu_{\rm L}/2\pi}{(\nu - \nu_0)^2 + (\Delta \nu_{\rm L}/2)^2}, 
\end{equation}
where $\Delta \nu_{\rm L} = 9.936\times10^7~\rm{Hz}$ and $\nu_0=2.466\times 10^{15}~\rm{Hz}$ are the natural broadening width 
and the central frequency of the Ly$\alpha$ line, respectively. 
The cross-section $\sigma_\nu$ of the \Lya scattering in the rest-frame of hydrogen atom is 
\begin{equation}
	\sigma_\nu = f_{12}\frac{\pi e^2}{m_{\rm e} c} \phi(\nu) = f_{12}\frac{\pi e^2}{m_{\rm e} c}\frac{\Delta \nu_{\rm L}/2\pi}{(\nu - \nu_0)^2 + (\Delta \nu_{\rm L}/2)^2}, 
\end{equation}
where $f_{12} = 0.4162$ is the \Lya oscillator strength, $m_{\rm e}$ is the mass of electron, and $c$  is the speed of light. 
When we consider the motion of an atom, the line center frequency shifts to $\nu_0(1+v_{||}/c)$ due to the Doppler shift in the laboratory frame, where $v_{||}$ denotes the velocity component of the atom parallel to the incident direction of a photon. 
Therefore, if the atoms have a thermal velocity distribution function (i.e., Maxwellian), the line profile in the laboratory frame is the well-known Voigt profile and can be obtained by summing up the Lorentzian profiles with various velocities; 
\begin{equation}
	\phi_{\rm H}(\nu) = \frac{\Delta \nu_{\rm L}}{2\pi}\int_{-\infty}^{\infty}dv_{||}\frac{(m_{\rm H}/2\pi k_{\rm B} T)^{1/2}\exp(-m_{\rm H}v_{||}^2/2k_{\rm B} T)}{(\nu-\nu_0-\nu_0 v_{||}/c)^2 + (\Delta \nu_{\rm L}/2)^2}, 
\end{equation}
where $T$ is the temperature, $k_{\rm B}$ is the Boltzmann constant and $m_{\rm H}$ is the neutral hydrogen mass. 
A normalized frequency, $x = (\nu - \nu_0)/\Delta \nu_{\rm D}$, is often used for the frequency distribution function, 
where $\Delta \nu_{\rm D} = (v_{\rm th}/c)\nu_0$ denotes the thermal Doppler broadening width, and $v_{\rm th} = \sqrt{2k_{\rm B}T/m_{\rm H}}$ corresponds to the thermal velocity dispersion. 
With these expressions, the \Lya scattering cross-section is described as 
\begin{equation}
	\sigma_{{\rm H}, x} = f_{12}\frac{\sqrt{\pi} e^2}{m_{\rm e}c\Delta \nu_{\rm D}}H(a,x), 
\end{equation}
where $a=\Delta \nu_{\rm L}/(2\Delta \nu_{\rm D})$ is the relative line width. 
$H(a,x)$ is the Voigt function defined as 
\begin{equation}
	H(a,x) = \frac{a}{\pi}\int_{-\infty}^{\infty}dy\frac{e^{-y^2}}{(x-y)^2+a^2}. 
\label{Voigt-function}
\end{equation}
The Voigt profile is composed of a Gaussian core and power-law wings. 
Although the Voigt function cannot be integrated analytically, \citet{Tasitsiomi06} has provided a useful analytical fitting formula for the Voigt function in the form
\begin{equation}
	H(a,x) \sim q\sqrt{\pi}+e^{-x^2}, 
\end{equation}
where  
\begin{eqnarray}
	\label{eq:H_fitting}
	q &=&  \left\{ \begin{array}{ll}
     	 \displaystyle{0} & {\rm for}~ z \le 0 \\ \nonumber
      	\left(1+\frac{21}{x^2}\right)\frac{a}{\pi(x^2+1)}P(z) & {\rm for}~ z > 0, 
    	\end{array}
    	\right. \\ \nonumber
	P(z) &=& 5.674z^4 -9.207z^3+4.421z^2+0.1117z, \\ \nonumber
	z &=& (x^2-0.855)/(x^2+3.42).
\end{eqnarray}
We adopt this formula to evaluate the \Lya scattering cross-section in our code. 
It is worth mentioning that the cross-section in the wing is typically $\sim 10^5$ times smaller than that at the line center. 
Hence, once a frequency is shifted to the wing part, the photon can easily escape from a system even if the system is quite optically thick at the line center frequency. 
 
If a photon is scattered by an atom in the direction parallel to the incident one, 
the frequency of the photon does not vary, because the scattering is coherent in the atom's rest frame.
However, if the scattering direction is not parallel, the frequency is shifted owing to the motion of the atom. 
This results in the partial frequency redistribution.
We denote the incident frequency of the photon in the laboratory frame by $\nu_{\rm in}$, its incoming direction by $\bm{d}$, the atom's microscopic velocity by $\bm{v}$, and the bulk (macroscopic) velocity of the fluid by $\bm{V}$. 
Using these expressions, the incident frequency in the atom rest-frame $\nu_{\rm atom}$ is written as
\begin{equation}
	\label{eq:nu_atom}
	\nu_{\rm atom} = \nu_{\rm in} \left(1-\frac{\bm{v}\cdot \bm{d}}{c} - \frac{\bm{V}\cdot \bm{d}}{c}\right). 
\end{equation}
After the scattering, the photon flies in a different direction $\bm{d'}$. 
Since the scattering is coherent in the rest-frame of the atom, the frequency of the scattered photon in the laboratory frame $\nu_{\rm out}$ turns out to be 
\begin{equation}
	\label{eq:nu_out}
	\nu_{\rm out} = \nu_{\rm atom} / \left(1-\frac{\bm{v}\cdot \bm{d'}}{c} - \frac{\bm{V}\cdot \bm{d'}}{c}\right). 
\end{equation}
It should be noted that the scattering direction is actually dependent on the phase function, 
which is determined by the excitation state and scattering frequency \citep[e.g., see][]{Tasitsiomi06}. 
However, for an optically thick medium, the anisotropy of the phase function is not significant 
for the emergent spectrum and/or the escape fraction \citep{LICORICE,Laursen+09,ART2}. 
Therefore, we assume a simple isotropic scattering in our code. 

Since the frequency shift is determined by the velocity component in the scattering direction, 
its probability depends on the shape of the velocity distribution function. 
To quantify the frequency shift in our code, we split the velocity components of the atoms into $v_{\bot}$ and $v_{||}$ that are respectively the perpendicular and parallel components with respect to $\bm d$. 
For the perpendicular component $v_{\bot}$, the velocity distribution function is a Maxwellian. 
Therefore, the distribution function of the perpendicular component normalized by the thermal velocity dispersion 
($u_{\bot} \equiv v_{\bot}/v_{\rm th}$) is given by 
\begin{equation}
	\label{eq:g_distri}
	g(u_{\bot}) = \frac{1}{\sqrt{\pi}} e^{-u_{\bot}^2}. 
\end{equation}
On the other hand, 
for a given incoming relative frequency $x$, the scattering cross section is  
the superposition of the Lorentz profiles shifted in terms of 
the parallel velocities of different atoms, which is described by the Voigt function (\ref{Voigt-function}). 
Hence, the probability that a photon is scattered by an atom with a certain parallel velocity 
$u_{||} (\equiv v_{||}/v_{\rm th})$ is given by
\begin{equation}
	\label{eq:f_distri}
	f(u_{||}) = \frac{a}{\pi H(a,x)}\frac{e^{-u_{||}^2}}{(x-u_{||})^2+a^2}.
\end{equation}
In the case of $|x| \ll 1$, photons are selectively scatted by atoms with $u_{||}\approx x$. 
As a result, the probability function $f(u_{||})$ exhibits a sharp peak at $u_{||}\approx x$. 
On the other hand, for large $|x|$, the number of atoms with $u_{||}\approx x$ exponentially decreases and therefore photons are scattered predominantly in the Lorentz wing.  
In this case, the distribution function behaves as a Gaussian distribution \citep[e.g., see Fig. 2 of][]{Laursen+09}. 

In the present code, we randomly generate the perpendicular components $u_{\bot}$ by the Box-Muller method, 
\begin{eqnarray}
	u_{\bot,1} = \sqrt{-\ln R_1}\cos(2\pi R_2) \nonumber \\
	u_{\bot,2} = \sqrt{-\ln R_1}\sin(2\pi R_2), 
\end{eqnarray}
where $R_{1}$ and $R_{2}$ are two univariates. 
The parallel component $u_{||}$ is randomly generated to obey the distribution function (\ref{eq:f_distri}) in the same way as in \citet{Zheng&Miralda02}.

\subsection{Ray-tracing in {\small\sf SEURAT}}\label{sec:RT}
The method for integrating optical depths in our code is fairly distinct from those in the previous Monte Carlo schemes of line transfer, since our ray-tracing algorithm is optimized for using SPH particles directly. 
Ray-tracing schemes of continuum radiative transfer optimized for SPH simulations have hitherto succeeded in handling a wide dynamic range by the Lagrangian description of SPH  \citep[e.g.,][]{Kessel&Burkert00, RSPH, TRAPHIC, SPHRAY, START}. 

\begin{figure}
	\begin{center}
		\includegraphics[width=8.5cm,clip]{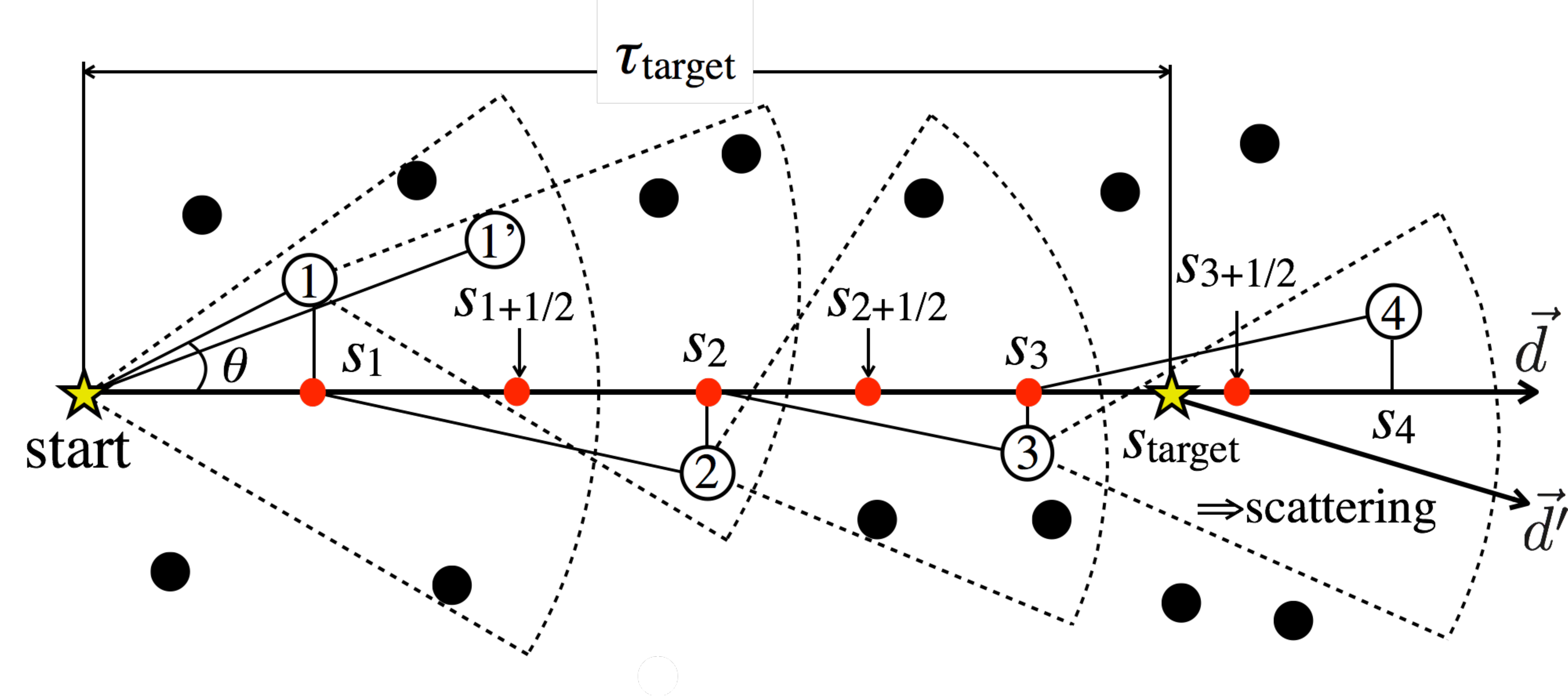}
	\end{center}
	\caption{
		Schematic illustration of the ray-tracing scheme in {\sf SEURAT}. 
		The filled circles and numbered open circles represent the SPH particles. 
		In the ray-tracing, we search the downstream particle which is the closest to the ray. 
		Grid points are determined by the projection of the downstream particles, 
		as shown by red dots, 
		and the optical depth is evaluated at intermediate points between two grid points. 
		We do not use particle 1' as a downstream particle since particle 1 is closer to the ray,
		although particle 1' has the smaller angle. 
		The dashed sectors represent the region where candidates for the next downstream particle are searched.  
		Note that the radii of the sectors are not necessarily equal to the SPH smoothing length (see \S \ref{sec:discussion}). 		
	}
	\label{fig:RT}
\end{figure}
The ray-tracing algorithm in {\small\sf SEURAT} is similar to that in \cite{RSPH} except that rays are not always cast towards SPH particles. 
Fig. \ref{fig:RT} shows the principle of the ray-tracing algorithm in {\small\sf SEURAT}. 
After determining the direction of the light ray from a radiation source, we search the  particle closest to the light ray (particle 1 in Fig. \ref{fig:RT}) from the list of neighbour particles of the source . 
Then we create a virtual grid point at the position where the perpendicular line intersects with the ray, and directly assign the physical quantities of the particle such as the density, the temperature and the velocity component ($\bm{d}\cdot \bm{v}$) to the grid point (grid point $s_1$ in Fig. \ref{fig:RT}). 
We define the next downstream grid point (grid point $s_2$ in Fig. \ref{fig:RT}) on the light ray in the same way except that we search from the list of neighbour particles of particle 1. 
The optical depth is then integrated with the following formula, 
\begin{equation}
	\tau_{i+1/2} = \tau_{i-1/2} + \Delta \tau_i, 
	\label{eq:tau}
\end{equation}
\begin{equation}
	 \Delta \tau_i = \sigma_{x_i} n(\bm{r_i}) \left(s_{i+1/2} - s_{i-1/2}\right), 
	\label{eq:dtau}
\end{equation}
where $\sigma_{x_i}$ is the scattering cross-section of a \Lya photon at a frequency $x_i$ and $n(\bm{r_i})$ is the number density at a particle position $\bm{r_i}$ that is obtained by the superposition of neighbouring SPH particles as
\begin{equation}
	\label{eq:rho_SPH}
	n(\bm{r_i}) = \sum_j m_j W(|\bm{r_i} - \bm{r_j}|, h_i)/m_{\rm H}, 
\end{equation}
where $m_j$, $h_i$, and $W$ respectively denote the mass of the $j$-th neighbour particle, the smoothing length of the $i$-th particle, and the kernel function. 
The accuracy of our ray-tracing scheme will be discussed in \S~\ref{sec:SEURAT-LICORICE}. 
The $s_{i\pm1/2}$ is the position defined as the intermediate point between the grid points $s_i$ and $s_{i\pm1}$ on the light ray. 
Here we assume that the physical quantities are constant between the back and front of the $i$-th particle.  
This formulation is employed to avoid the mixing of the physical quantities between adjacent SPH particles. 
Especially for velocities, if adjacent SPH particles move with the similar speed in opposite directions, 
the mixing results in canceling out the velocities and underestimating the $\bm{d}\cdot \bm{v}$ term artificially. 
Repeating calculations described by Eqs.(\ref{eq:tau}) and (\ref{eq:dtau}) along the light ray, we integrate the optical depth while $\tau_{i+1/2} < \tau_{\rm target}$. 
If the condition of $\tau_{i+1/2} >  \tau_{\rm target}$ is satisfied, we determine the position $s_{\rm target}$ where a scattering occurs by using the physical quantities of $i$-th particle (in Fig. \ref{fig:RT}, the $i$-th particle corresponds to the particle 3), i.e., 
\begin{equation}
	\tau_{\rm target} = \tau_{i-1/2} + \sigma_{x_i} n(\bm{r_i}) \left(s_{\rm target} - s_{i-1/2}\right). 
\end{equation}
On scattering at $s_{\rm target}$, we stochastically choose a new direction $\bm{d'}$ to which the scattered photon packet travels, and let the packet propagate until a subsequent scattering occurs. 
If there is no neighbour downstream particle during the ray-tracing, the position is regarded as a boundary of the system, and we assume that the photon packet escapes from the system.

It is worth mentioning that the scattering condition $\tau_{i+1/2} >  \tau_{\rm target}$ is often satisfied before the photon packet renews the downstream particle, since the local optical depth is very large. 
In this case, to ensure continuity of the physical quantities (e.g., velocities) before and after the scattering, the same particle $i$ is used as the origin of the subsequent ray-tracing rather than the particle closest to the packet. 
Consequently, multiple scattering events frequently occur around a single particle and the photon packet tends to remain within its smoothing kernel length. 
We will discuss how this behavior affects to the SPH-based ray-tracing in \S~\ref{sec:discussion}.

\subsection{Dust absorption and scattering}\label{sec:dust}
The absorption and scattering by dust grains make a significant effect on the \Lya radiative transfer in metal-enriched media. 
The optical depth for the dust scattering and absorption can be written: 
\begin{equation}
	\label{eq:tau_dust}
	d\tau_{{\rm d},x} = (Q_{{\rm s},x} + Q_{{\rm a},x}) n_{\rm d} \pi r_{\rm d}^2 ds 
	\equiv d\tau_{{\rm d, s},x} + d\tau_{{\rm d, a},x},  
\end{equation}
with 
\begin{equation}
	n_{\rm d} = n_{\rm H}\left(\frac{m_{\rm H}}{m_{\rm d}}\right) f_{\rm d}.
\end{equation}
where $r_{\rm d}$ is the dust grain radius, $Q_{s,x}$ and $Q_{a,x}$ respectively denote the  scattering and absorption efficiencies
(so-called $Q$-value) at a frequency $x$, and $n_{\rm d}$ is the dust density which is determined by the dust grain mass $m_{\rm d}$ and the dust-to-gas mass ratio $f_{\rm d}$. 
In the range of UV frequencies, both $Q_{{\rm s},x}$ and $Q_{{\rm a},x}$ are $\sim1$ \citep{Verhamme+06}, i.e., they are independent of the photon frequency. 
Then, the total optical depth $d\tau_{{\rm tot},x}$ including contributions by hydrogen and dust is given by 
\begin{equation}
	\label{eq:tau_w_dust}
	d\tau_{{\rm tot},x} 
	=(n_{\rm H_{I}} \sigma_{{\rm H}, x} + n_{\rm d} \sigma_{\rm d}) ds  
	\equiv d\tau_{{\rm H},x} + d\tau_{{\rm d, s}} + d\tau_{{\rm d, a}},  
\end{equation}
where $d\tau_{{\rm H},x}$, $d\tau_{{\rm d, s}}$ and $d\tau_{{\rm d, a}}$ correspond to the optical depths for the hydrogen scattering, the dust scattering, and the dust absorption respectively. 

The probability that \Lya photons are scattered by hydrogen is given by 
\begin{equation}
	\label{eq:pHx}
	p_{{\rm H},x} = \frac{n_{\rm H_{I}} \sigma_{{\rm H}, x}}{n_{\rm H_{I}} \sigma_{{\rm H}, x} + n_{\rm d} \sigma_{\rm d}}. 
\end{equation}
We generate a uniform random number  $\xi'$ between 0 and 1, and
let the photon packet interact with hydrogen if $\xi' < p_{{\rm H},x}$ and otherwise interact with dust.  

We invoke two different methods 
to assess the absorption by dust; 
one is the ``stochastic elimination method'' and  the other is the ``flux attenuation method''.
When a photon packet interacts with dust, the fraction of scattering is given by
the scattering albedo, $\varpi = Q_{\rm s}/(Q_{\rm s} + Q_{\rm a})$. 
In the stochastic elimination method, we generate another uniform random number $\xi''$, 
and eliminate the photon packet if $\xi'' < 1-\varpi$. 
On the other hand, in the flux attenuation method, we reduce the number of photons in a photon packet in accordance 
with an attenuation factor $\exp(-\tau_{\rm d,a})$ during the journey. 
To avoid double counting in the latter case, we neglect the third term $d\tau_{d, \rm a}$ 
in Eq.~(\ref{eq:tau_w_dust}) when we integrate $d\tau_{{\rm tot},x}$ up to $\tau_{\rm target}$. 

\section{Tests of the code}
In this section, we present several test calculations to demonstrate the validity of the code. 
Throughout the tests, we model the systems with $64^{3}$ SPH particles, 
and pursue the propagation of $10^{5}$ photon packets emitted at the \Lya line center frequency. 
We note that it is unsuitable to define a geometrically sharp boundary of a system in the SPH formalism, 
since the gas density fields associated to the SPH particles are smoothed with the kernel function, 
even though a simple geometry such as a spherical cloud or a slab is assumed. 
Thus, for the test calculations, we regard photons beyond an assumed boundary as escaped ones.  
Note that previous Monte Carlo codes often adopted the core-skipping method which artificially avoids the significant number of the scattering events that happen in the core of the line and reduces the computational time \citep{Ahn+02}. 
However, we do not use the core-skipping scheme in this paper, since accelerating the computation is not our aim for now. 
We focus here on proving the validity of the meshfree Monte Carlo \Lya radiation transfer calculation.


\subsection{Static homogeneous spherical cloud}
For a static homogeneous spherical cloud, the analytical formula of the emergent spectrum was derived by \citet{Dijkstra+06} as follows, 
\begin{equation}
	\label{eq:Dijkstra}
	J(x) = \frac{\sqrt{\pi}}{\sqrt{24}a\tau_0}\left\{ \frac{x^2}{1+ {\rm cosh} \left[\sqrt{2\pi^3/27}(|x^3|/a \tau_0)\right]} \right\}, 
\end{equation}
where $\tau_0$ denotes the line center optical depth from the center to the boundary of the cloud. 
We test three cases of optical depth, $\tau_0 = 10^4$, $10^5$ and $10^6$, 
for a spherical cloud with the gas temperature of 10~K. 
SPH particles are randomly distributed in the cloud. 
The source of \Lya photons is located at the center of the cloud. 
\begin{figure}
	\begin{center}
		\includegraphics[width=9cm,clip]{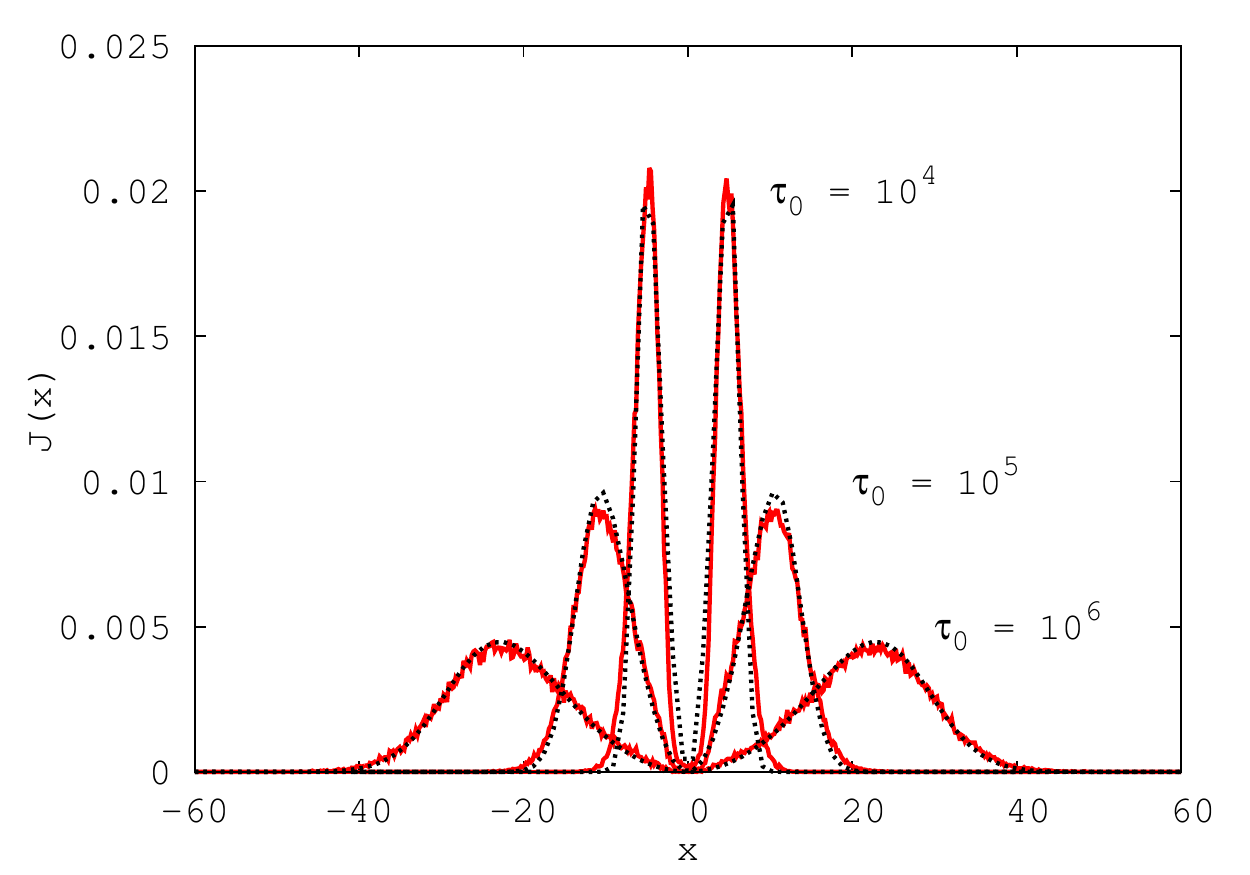}
	\end{center}
	\caption{
		Emergent spectra from a static homogeneous spherical cloud for three different optical depths. 
		The line center optical depth of the cloud is shown by the attached number. 
		The temperature of the clouds is assumed to be 10~K, and the photon source is located at the center of the cloud. 
		Solid lines represent the results of the numerical calculations, while dotted lines denote the analytical solutions derived by \citet{Dijkstra+06}. 
	}
	\label{fig:test1}
\end{figure}
Fig. \ref{fig:test1} shows the comparison between the analytical solutions of \citet{Dijkstra+06} and our numerical calculations. 
We can see in Fig. \ref{fig:test1} that the numerically calculated emergent spectra (solid lines) are in good agreement with the analytical solutions (dotted lines). 
Strictly speaking, the agreement between the numerical and analytical emergent spectra tend to be better at large optical depth, 
since the analytic formula is valid for extremely optically thick media, say, $a\tau_0 \gtrsim 10^3$ \citep{Neufeld90}. 
In this test, $T=10$~K corresponds to $a\sim 1.5\times 10^{-2}$. Thus, the agreement for the corresponding $\tau_0 = 10^4$ spectrum is slightly worse when compared to more optically thick cases. 
The resultant spectra reproduce the double-peaked shapes, and the positions of the peak frequencies move outward with increasing the optical depth of the system. 
This is because only \Lya photons with frequencies far from the line center can escape from the clouds as $\tau_0$ increases. 

\subsection{Expanding homogeneous spherical cloud}\label{sec:expanding}
Next we calculate the emergent \Lya spectrum from an expanding spherical cloud.
For such a moving medium, there is no analytical solution. 
However, for a homogeneous expanding/infalling cloud, it is comparatively easy to formulate a physical interpretation of the emergent spectrum. 
Hence, this test is useful to demonstrate the validity of the code, and has been used in previous works
\citep[e.g.,][]{Zheng&Miralda02,Verhamme+06,Dijkstra+06,LICORICE,Laursen+09,ART2}. 
We consider a uniform spherical cloud with a Hubble-like velocity field 
\begin{equation}
       v(r) = v_{\rm max} \left( \frac{r}{r_{\rm max}} \right), 
\end{equation}
where $v_{\rm max}$ is the radial velocity at the edge of the cloud, 
and $r_{\rm max}$ is the radius of the cloud. 
We assume a gas temperature of 20,000~K 
and $v_{\rm max} = 200~{\rm km~s^{-1}}$ \citep[e.g.,][]{Zheng&Miralda02}. 
The neutral hydrogen column densities are set to be $N_{\rm H}=2\times10^{18}~\per{cm}{2}$, $2\times10^{19}~\per{cm}{2}$ and $2\times10^{20}~\per{cm}{2}$. 
They correspond to $\tau_0 = 8.3\times10^{4}$, $\tau_0 = 8.3\times10^{5}$ and $\tau_0 = 8.3\times10^{6}$, respectively. 
The source is located at the center of cloud. 

\begin{figure}
	\begin{center}
		\includegraphics[width=9cm,clip]{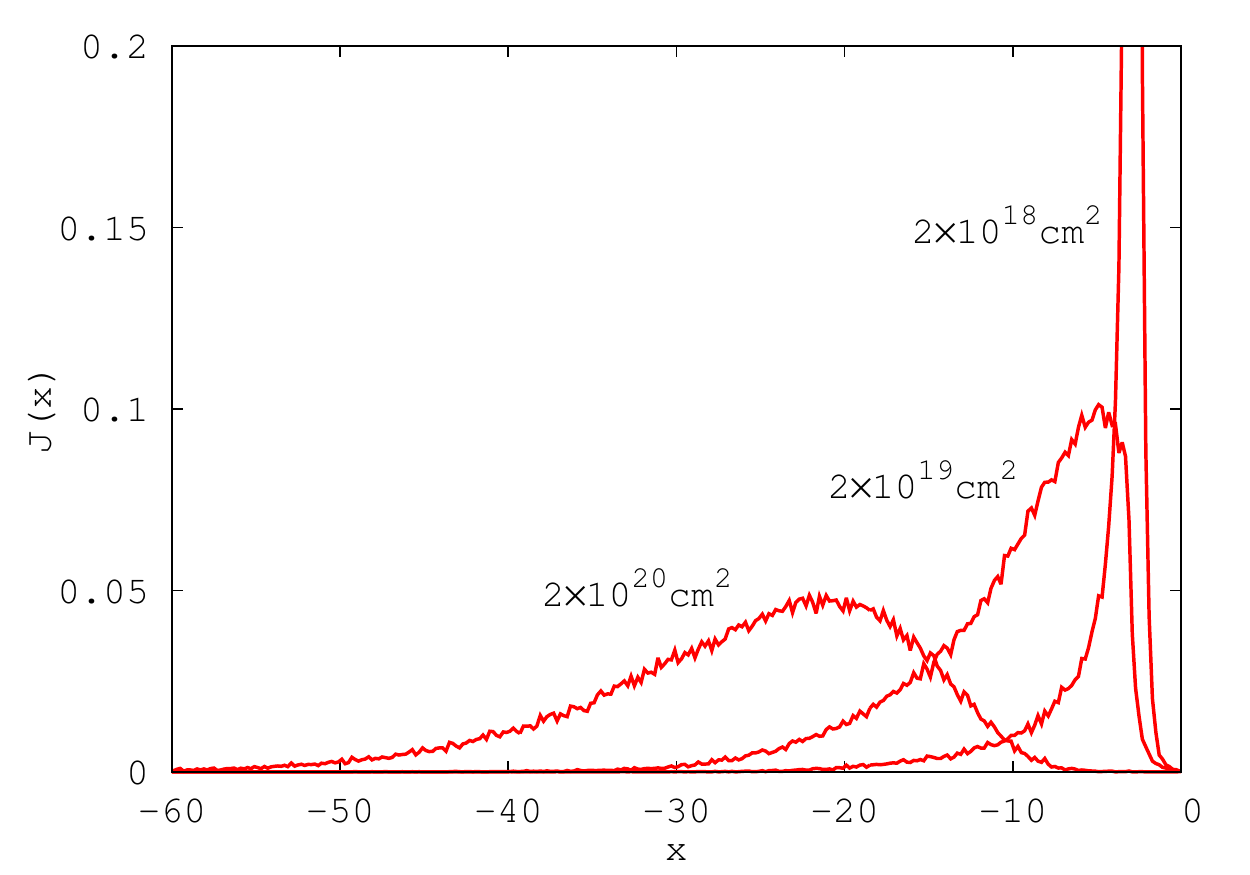}
	\end{center}
	\caption{ 
		Emergent spectra from an expanding spherical cloud. 
		The temperature of the cloud is assumed to be 20,000~K, and 
		three different column densities of hydrogen are assumed. 
		The radial velocity is proportional to radius $r$ with the velocity of  200~$\rm km~s^{-1}$ at the edge. 
	}
	\label{fig:test2}
\end{figure}

The emergent spectra are presented in Fig. \ref{fig:test2}. 
The bulk expanding motion of the medium shifts the resonant frequency blueward in the laboratory frame (see \S\ref{sec:physics}). 
Hence, the blueward photons efficiently interact with hydrogen as they travel toward outer regions of the cloud.
As a result, the emergent spectra acquire asymmetric shapes with an emission peak in the redward frequencies.
The resultant emergent spectra are in good agreement with those from previous Monte Carlo codes \citep[e.g., see Fig. 2 of][]{LICORICE}. 
Here, we do not present the infalling case, because the emergent spectrum becomes basically in a mirrored shape 
with respect to $x=0$. 

\subsection{Static dusty slab} 
The escape fraction of \Lya photons from a dusty slab is a standard test, 
since it also can be solved analytically. \citet{Neufeld90} derived the analytical formula 
of  the escape fraction $f_{\rm esc}$ as
\begin{equation}
	\label{eq:fesc}
	f_{\rm esc} = \frac{1}{\cosh \left[\zeta' (a\tau_0)^{1/3}\tau_{\rm d,a}\right]},
\end{equation}
where $\zeta' = \sqrt{3}/(\zeta \pi^{5/12})$ with $\zeta = 0.525$, a fitting parameter. 
This formula is valid in the case of an optically thick medium with $(a\tau_0)^{1/3} \gg \tau_{\rm d,a}$. 
For the test calculation, we assume spherical silicate dust grains with a radius $r_{\rm d}$ of $2.4\times10^{-2}~\rm \mu m$ and 
solid material density of 3~$\rm g~cm^{-3}$ \citep{Draine&Lee84} so that the dust opacity is equal to that in \citet{Verhamme+06}. 
Note that the dust opacity under this assumption of single-sized grains is equivalent 
to that for the dust distributions by MRN \citep{Mathis+77} given the geometrical cross-section
in the range from $5.8\times 10^{-4}~{\rm \mu m}$ to 1~$\rm \mu m$ \citep{Yajima+17}.
%

\begin{figure}
	\begin{center}
		\includegraphics[width=9cm,clip]{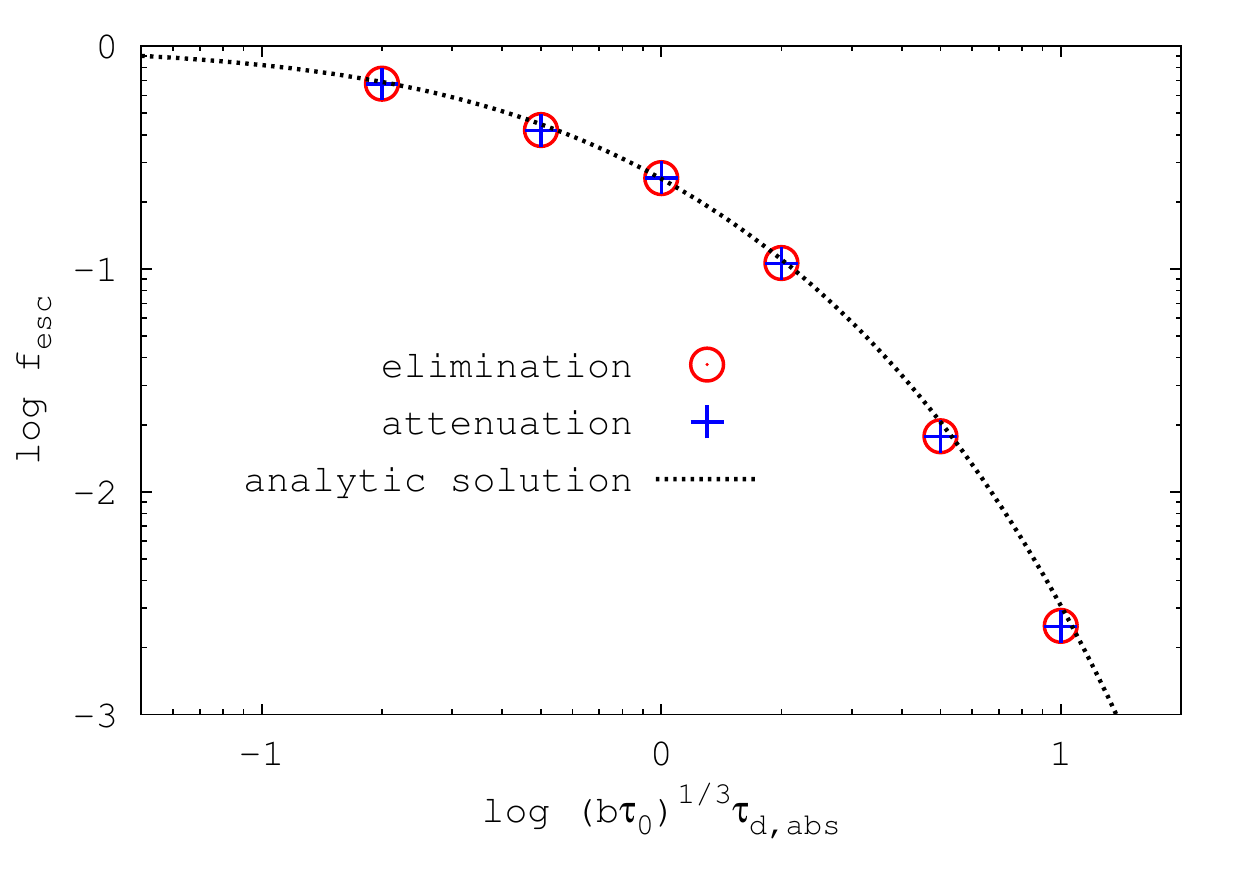}
	\end{center}
	\caption{ 
		The escape fraction from a dusty slab as a function of $(b\tau_0)^{1/3} \tau_{\rm d,a}$. 
		Red circles and blue crosses are respectively the results in 
		the ``stochastic elimination of photon packets'' method and in the ``attenuation of flux'' method 
		(see section 2.3 for the detail). 
		The dotted line represents the analytical solution obtained by \citet{Neufeld90}. 
	}
	\label{fig:fesc}
\end{figure}

Fig. \ref{fig:fesc} shows the resultant escape fraction. 
We see in the figure that the numerically obtained escape fraction is concordant with the analytical solution, regardless of 
the adopted method for the dust absorption.

\section{Adaptivity of {\sf SEURAT}}\label{sec:discussion}

In the meshfree radiative transfer adaptive for SPH, we should often treat the propagation of \Lya photons
in a local region having a very high optical depth or a large density gradient. 
Here, we describe the contrivance incorporated in {\small\sf SEURAT}, which ensures the high adaptivity for media with high density contrast.

\subsection{High optical depth regions}
As explained in \S \ref{sec:RT},  photon packets are judged to escape from the system if the packets do not find a downstream particle. 
Therefore, an appropriate neighbour search is required. 
Otherwise, we wrongly judge the escape of a photon packet and 
it may impact the resultant emergent spectra and/or escape fraction. 
This point is critical for a meshfree Monte Carlo based \Lya radiative transfer scheme, 
but has never been a concern for mesh-based schemes, since downstream grid cells are always recognized. 

\begin{figure}
	\begin{center}
		\includegraphics[width=8cm,clip]{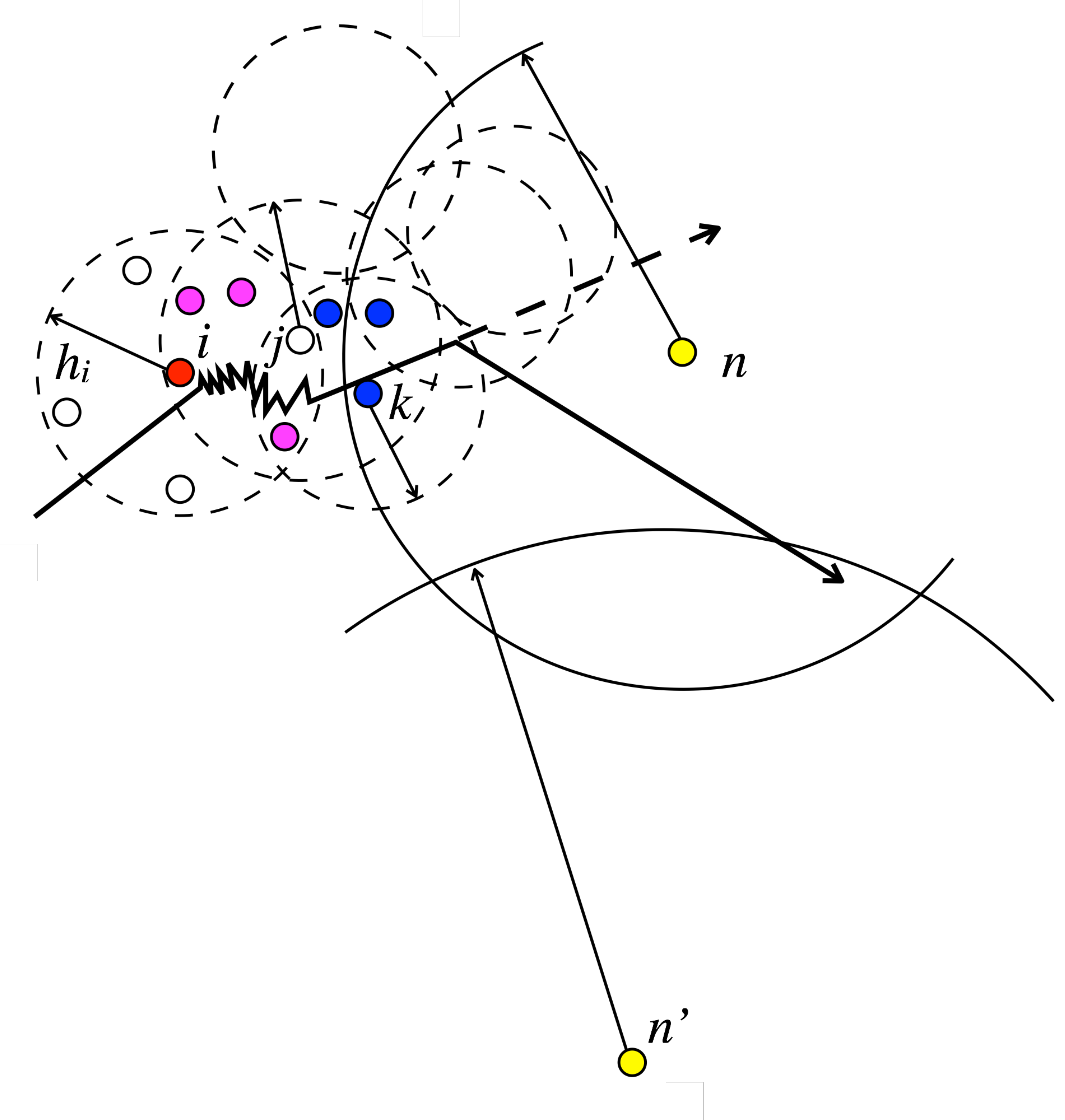}
	\end{center}
	\caption{
		Conceptual figure for the constitution of the neighbour list. 
		Red and open circles represent the $i$-th particle and its neighbour particles, respectively. 
		Blue circles are the neighbour particles of the $j$-th particle, but not of the $i$-th particle. 
		Magenta depicts the particles belonging to the neighbour lists of both $i$-th and $j$-th particles. 
		Dashed line denotes the smoothing kernel length of each SPH particle. 
		We suppose these particles are in a high-density region and
		another particle (yellow circle) is located in a low-density region. 
		The smoothing kernel length of the yellow particle is plotted by a solid line. 
		A thick solid line represents the trajectory of the photon packet. 
	}
	\label{fig:neighbour}
\end{figure}
\begin{figure*}
	\begin{center}	

	\begin{tabular}{ccc}
	
	\begin{minipage}{0.33\hsize}	
		\begin{center}
			\includegraphics[width=6cm,clip]{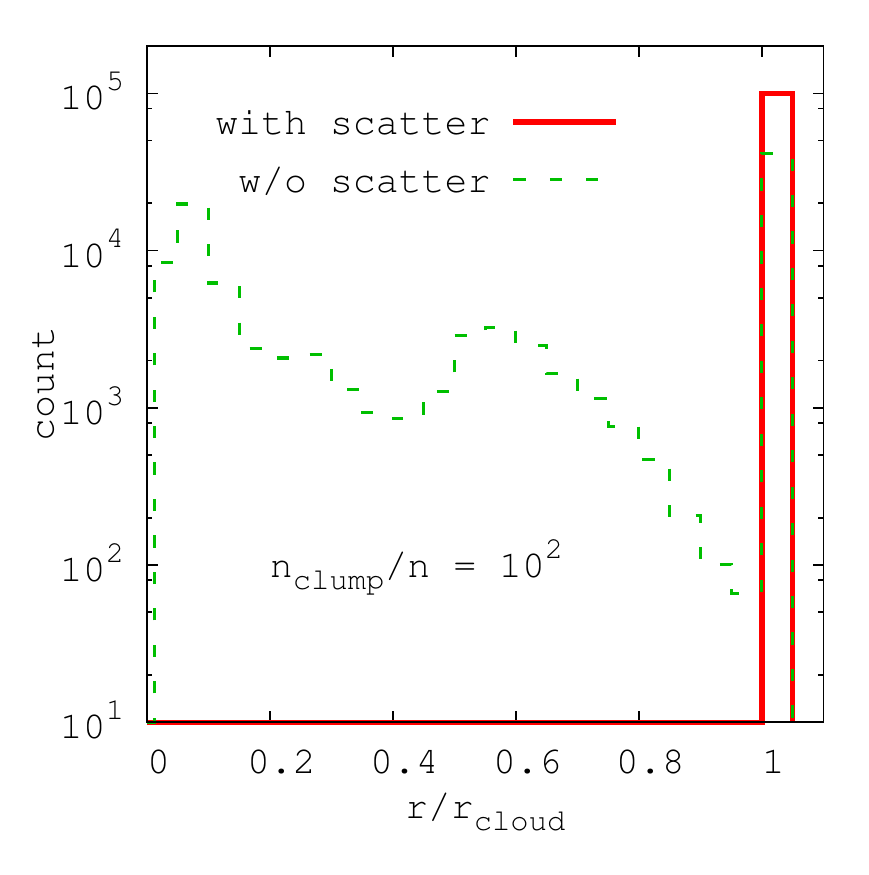}
		\end{center}
	\end{minipage}	
	\begin{minipage}{0.33\hsize}	
		\begin{center}
			\includegraphics[width=6cm,clip]{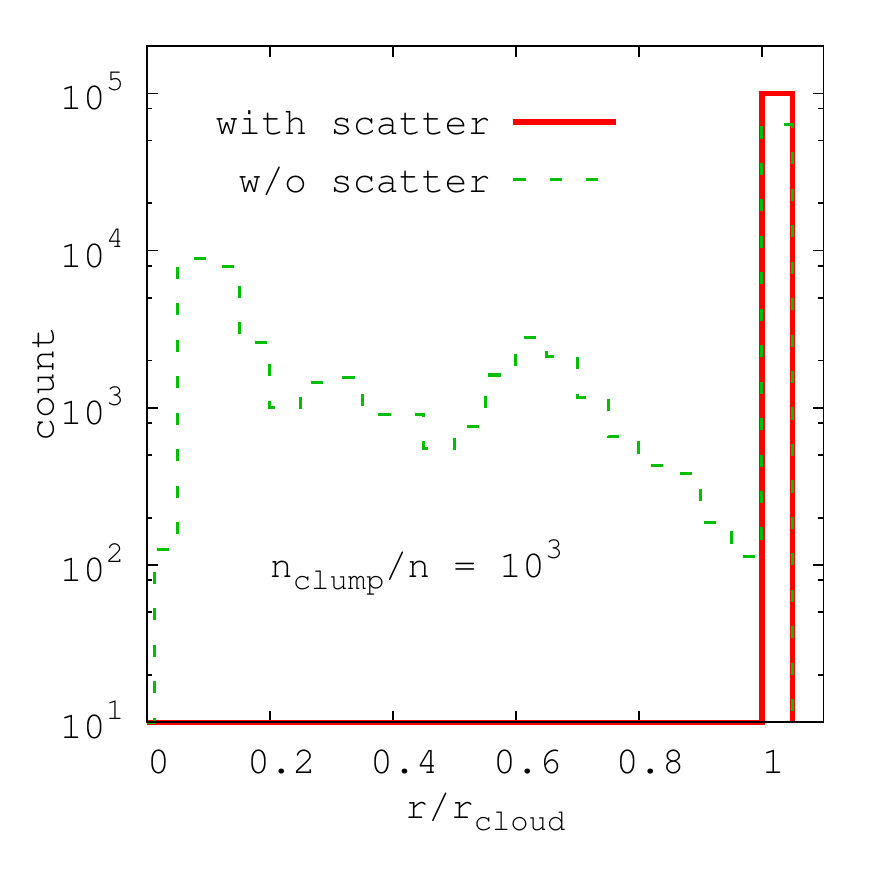}
		\end{center}
	\end{minipage}		
	\begin{minipage}{0.33\hsize}	
		\begin{center}
			\includegraphics[width=6cm,clip]{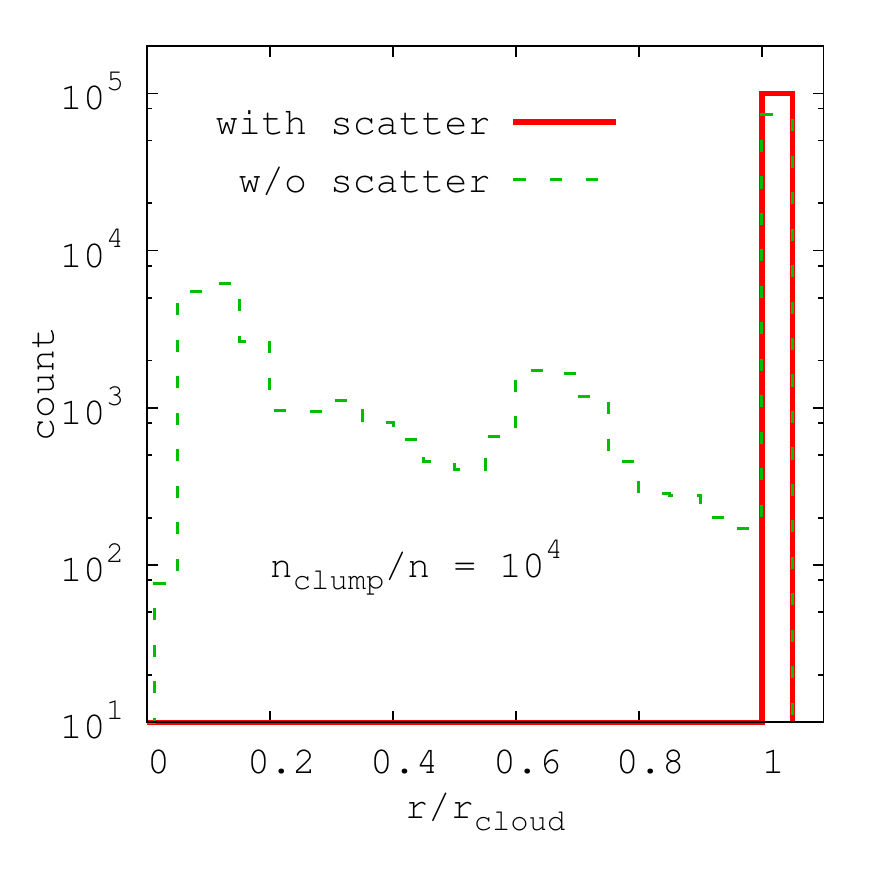}
		\end{center}
	\end{minipage}	
	
	\end{tabular}
	
	\end{center}
 
	\caption{Distribution of the positions at which photon packets escape. 
		From left to right, the panels show the results for the density contrast of $n_{\rm clump}/n_{\rm f}=10^2$, $10^3$ or $10^4$, respectively. 
		The vertical axis denotes the number count of the escape positions, 
		and the horizontal axis is the distance from the center of the cloud normalized by the cloud radius $r_{\rm cloud}$. 
		In each panel, a red solid line represents the results employing the scatter list in addition 
		to the ``neighbour of the neighbour'' list, while a green dashed line is the results dispensing with the scatter list. 
	}
	\label{fig:RT_compare}
\end{figure*}
Fig. \ref{fig:neighbour} shows a schematic view of the latent problems in the neighbour search procedure. 
In the meshfree Monte Carlo technique, the photon packets fly searching a neigbour particle. 
As a result, a ray-tracing algorithm sometimes fails if one uses the ``gather'' neighbour lists that are composed of the particles within the smoothing kernel length. 
This problem frequently arises, for instance, where the local optical depth is very large,  
since many scatterings occur around a single particle and the photon packet moves around the particle. 
When a scattering occurs at the boundary of the optically thick region, no downstream neighbour particle is found and then the packet is incorrectly labelled as an escaped packet
 (see the particle $i$ in Fig. \ref{fig:neighbour}). 
To circumvent this problem, we assign a larger number of neighbour particles 
by making the searching radius larger than the smoothing length $h_i$.
In addition, we extend the downstream particle search 
up to the ``neighbours of each neighbour particle'', if no downstream particle can be found from the neighbour list. 
We checked that this recipe successfully avoids false escapes in a simple uniform optically thick system. 
As a result, we correctly solved the \Lya radiative transfer in the standard test calculations 
as already presented in the previous section (Figs. \ref{fig:test1}, \ref{fig:test2} and \ref{fig:fesc}).
Note that, from the viewpoint of the computational cost, we avoid the naive double loop processing 
to track the neighbour list of the neighbour particles, since the two particles 
usually share some particles in their neighbour lists (depicted by magenta circles in Fig. \ref{fig:neighbour}). 
Instead, we construct the neighbour lists that contains only the ``neighbours of the neighbours'' 
(blue circles in Fig. \ref{fig:neighbour}).

\subsection{Large density gradient regions}

Moreover, incorrect escape events are highly probable when the local density gradient is quite large. 
Let us consider the particle $k$ in Fig. \ref{fig:neighbour} at the surface of a dense region surrounded by a low-density ambient gas and a photon packet propagating outward from the region. 
Since the ``gather neighbour list'' of the particle $k$ is mainly composed of particles in the dense region, 
we may not find any downstream particle of the particle $k$ even if we trace the ``neighbours of neighbours''. 
As a recipe coping with this situation, we construct the ``scatter list'' for the particle $k$, i.e., the list of particles which includes the particle $k$ in their gather neighbour list \citep{Hernquist&Katz89}, and use it instead of the gather neighbour list of the particle $k$. 
The gather lists of particles in the low density regions (particle $n$ in Fig. \ref{fig:neighbour}) often contains the particle $k$. 
Hence, we successfully find downstream particles for the particle $k$ from the scatter list. 
Therefore, we use the scatter list as the second fail-safe against false escapers.

\subsection{Highly inhomogeneous media}

To demonstrate the validity of the neighbour search scheme, we solve the \Lya radiative transfer in a highly inhomogeneous spherical cloud. 
In this test, we randomly distribute 16 clumps in a uniform spherical cloud. 
The total SPH number is $32^3$, and each clump consists of $32^2$ particles. 
We assume three models for the density of the clumps; $n_{\rm clump}/n_{\rm f} =10^2, 10^3$ and $10^4$, where $n_{\rm clump}$ and $n_{\rm f}$ denote the gas density for a clump and for the field, respectively. 
The clump size is comparable to the average separation between field particles 
in the model of $n_{\rm clump}/n_{\rm f} =10^3$. 
The source is located at the center of the cloud and the number of photon packets is set to be $10^5$. 
Fig. \ref{fig:RT_compare} shows the positions at which each photon packet is judged to escape.
As clearly seen in this figure, if we take only the ``neighbour of neighbour'' (green dashed line), 
photon packets escape before they reach the edge of the cloud. 
This is because, owing to the high density contrast, the neighbour lists of the SPH particles in the clumps 
hardly involve any SPH particles residing in the low-density field. 
As a result, photon packets in the high-density regions cannot find a downstream particle 
and therefore mistakenly label their positions as the boundary of the system. 
Thus, the ``neighbours of neighbours'' method is obviously insufficient to avoid incorrect escape of the photon packets. 
On the other hand, if we take the scatter list into account in the neighbour search procedure, 
we correctly solve the travel of photon packets (red solid line). 
As shown in the panels, all photon packets correctly recognize the edge of the system, 
regardless of the level of the density contrast.  
We emphasize that the scheme is valid even when the clump size is smaller than the typical separation 
of the field particles. 
Hence, we conclude that the neighbour search scheme in conjunction with the scatter list is indispensable to solve the meshfree Monte Carlo based radiative transfer.

\subsection{Comparison between {\small\sf SEURAT} and {\small\sf LICORICE}} \label{sec:SEURAT-LICORICE}
In this section, we compare the transfer in the Ly$\alpha$ lines with {\small\sf SEURAT} to that with {\small\sf LICORICE} which is a mesh-based RT code \citep{LICORICE}.

\subsubsection{Model galaxy}
\begin{figure}
	\begin{center}			
	 	\includegraphics[width=9.2cm,clip]{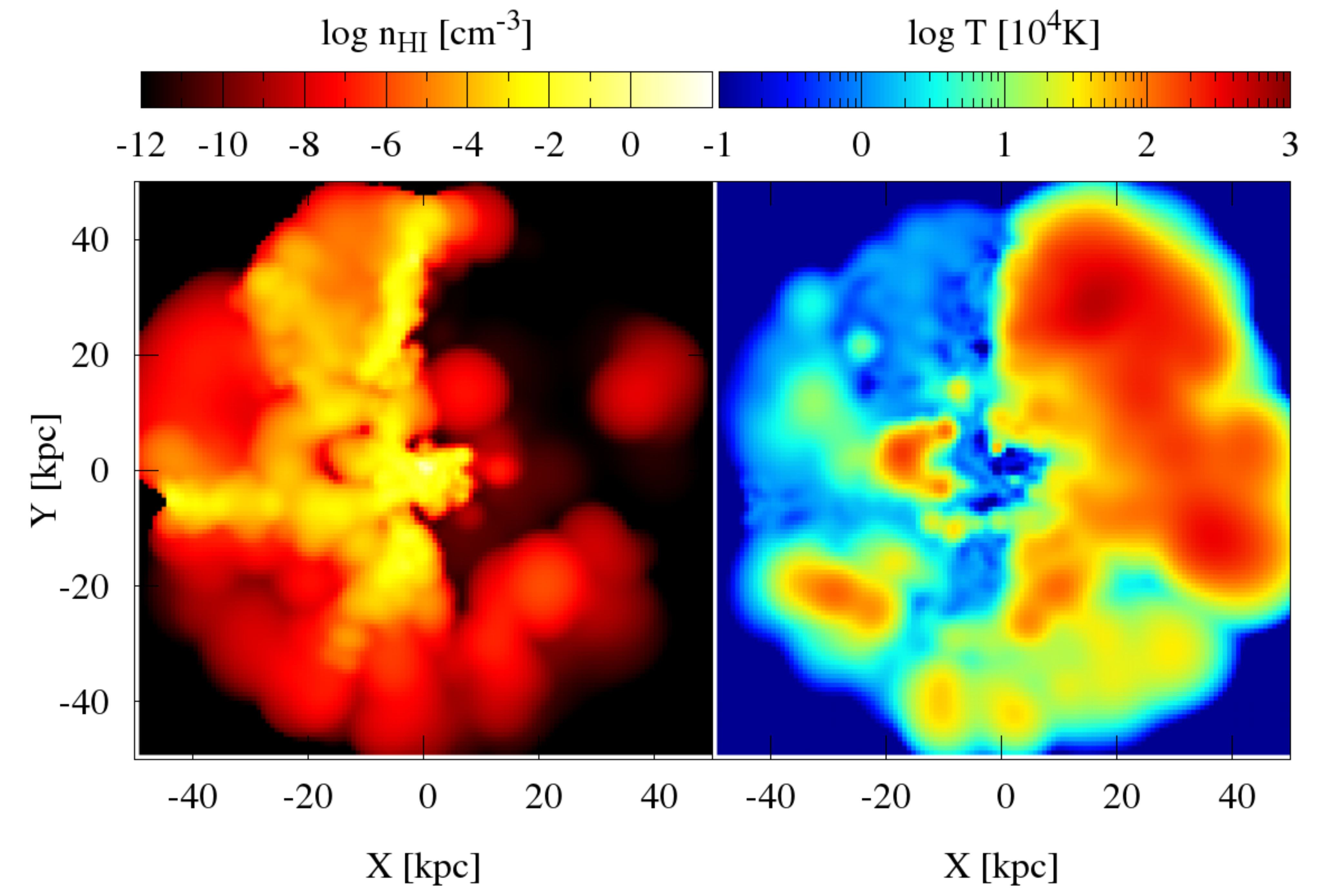}
	\end{center}
	\caption{Distributions of neutral hydrogen number density (left-hand panel) and temperature (right-hand panel) for a model galaxy. 
	}
	\label{fig:galaxy}	
\end{figure}

For this comparison, we use a model galaxy obtained by a cosmological radiation SPH simulation \citep[see][for the details of the simulation]{Hasegawa&Semelin13}. 
The model galaxy has a halo mass of $\sim 5.7\times 10^{10}~\solmass$ at $z\sim 6.1$ and is composed of $\sim 4.0\times 10^4$ SPH particles. The minimum smoothing length is $\sim 0.15$~kpc in the simulation. 
In {\small\sf LICORICE}, we generate a uniform grid with $512^3$ cells of size $\sim 0.20$~kpc, 
to resolve the minimum smoothing length of the SPH simulation. 
The radiative transfer of ionizing photons is solved coupled with hydrodynamics,
assuming the Case B recombination.
The resultant distributions of neutral hydrogen and temperature are shown in Fig. \ref{fig:galaxy}. 
Here, we do not incorporate dust extinction. 
As for the \Lya emissivity, we consider the recombination and the collisional excitation of hydrogen atoms.
Then, the local \Lya emissivity is respectively given by  
\begin{eqnarray}
	\epsilon_\alpha^{\rm rec} &=& f_\alpha \alpha_{\rm B} h \nu_\alpha n_{\rm H_{II}}  n_{\rm e} \\
	\epsilon_\alpha^{\rm col} &=& C_{{\rm Ly}\alpha} n_{\rm H_I}  n_{\rm e}, 
\end{eqnarray}
where $f_{\alpha}$ is the average number of the \Lya photons generated via the Case B recombination,
\citep[$f_{\alpha} = 0.68$,][]{Osterbrock&Ferland06}, $\alpha_{\rm B}$ is the Case B recombination coefficient \citep{Hui&Gnedin97}, $C_{{\rm Ly}\alpha}$ is the collisional excitation coefficient, $C_{{\rm Ly}\alpha} = 3.7\times 10^{-17} \exp(-h\nu_\alpha/k_{\rm B}T)T^{-1/2}~{\rm erg~s^{-1}~cm^3}$ \citep{Osterbrock&Ferland06}. 
We cast $10^5$ photon packets.

\subsubsection{Emergent spectrum}

\begin{figure}
	\begin{center}	
	\includegraphics[width=9cm,clip]{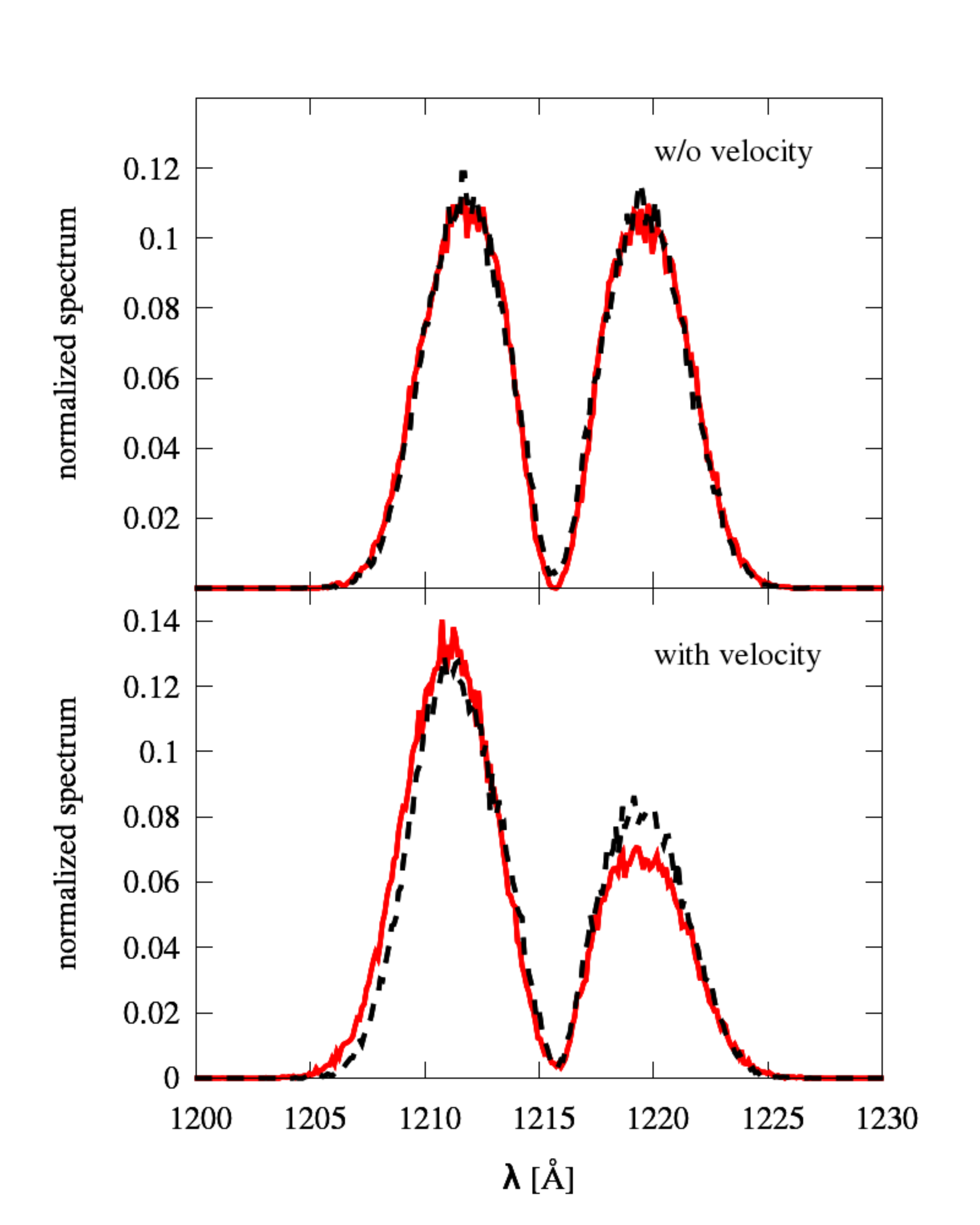}
	\end{center}
 	\caption{Emergent spectra from a model galaxy. 
		The horizontal axis is the wavelength and the vertical axis denotes the number count of photon packets in each wavelength bin. 
		A solid curve represents the result with {\small\sf SEURAT}, while a dashed curve does the result with {\small\sf LICORICE}. 
		The spectra are normalized such that the integration of the spectrum over frequencies should be unity. 
		In upper panel, the velocity structure is disregarded, while in the lower panel the velocity structure is included. 
	}
	\label{fig:SEURAT_LICORICE}
\end{figure}

\begin{figure}
	\begin{center}	
	\begin{tabular}{c}
	\begin{minipage}{0.99\hsize}	
		\begin{center}
			\includegraphics[width=8cm,clip]{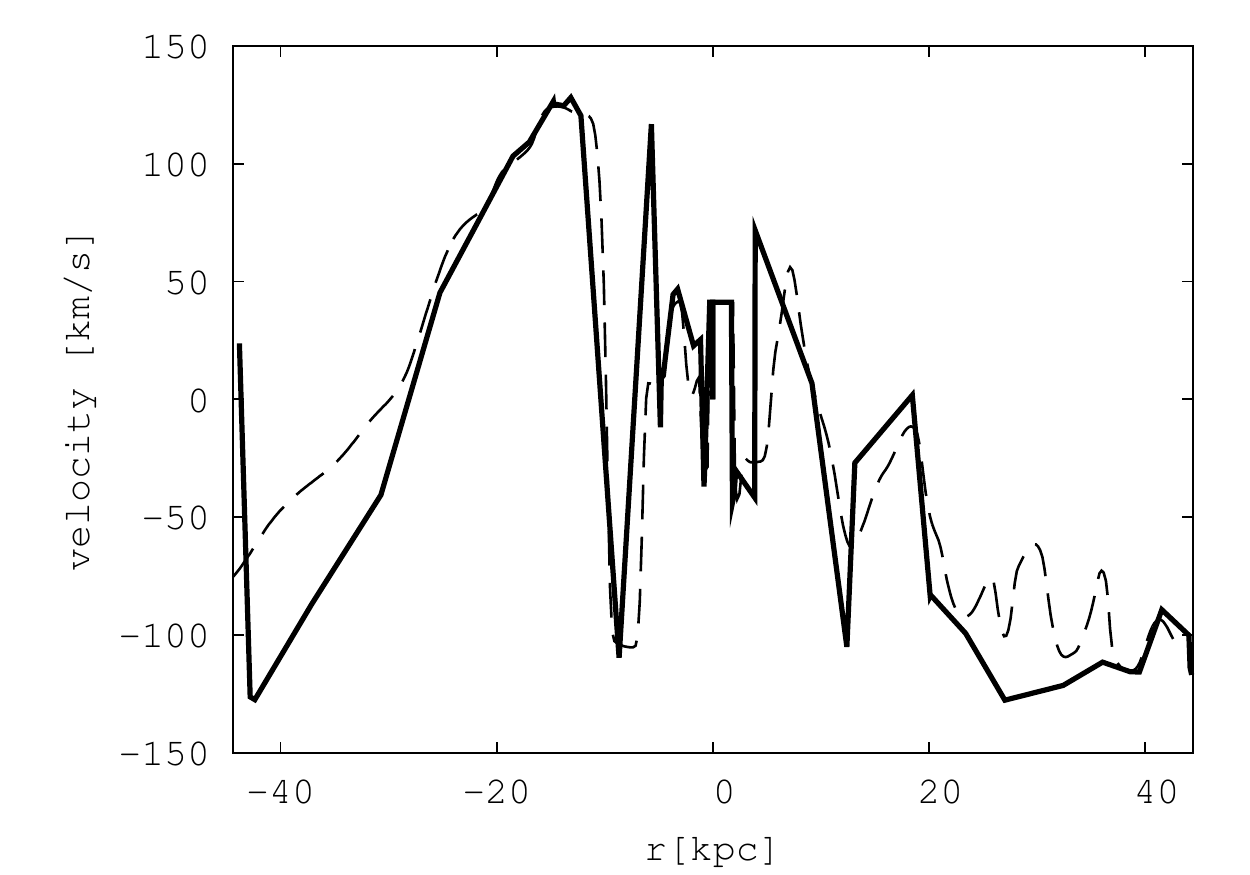}
		\end{center}
	\end{minipage}	
	\\
	\begin{minipage}{0.99\hsize}	
		\begin{center}
			\includegraphics[width=8cm,clip]{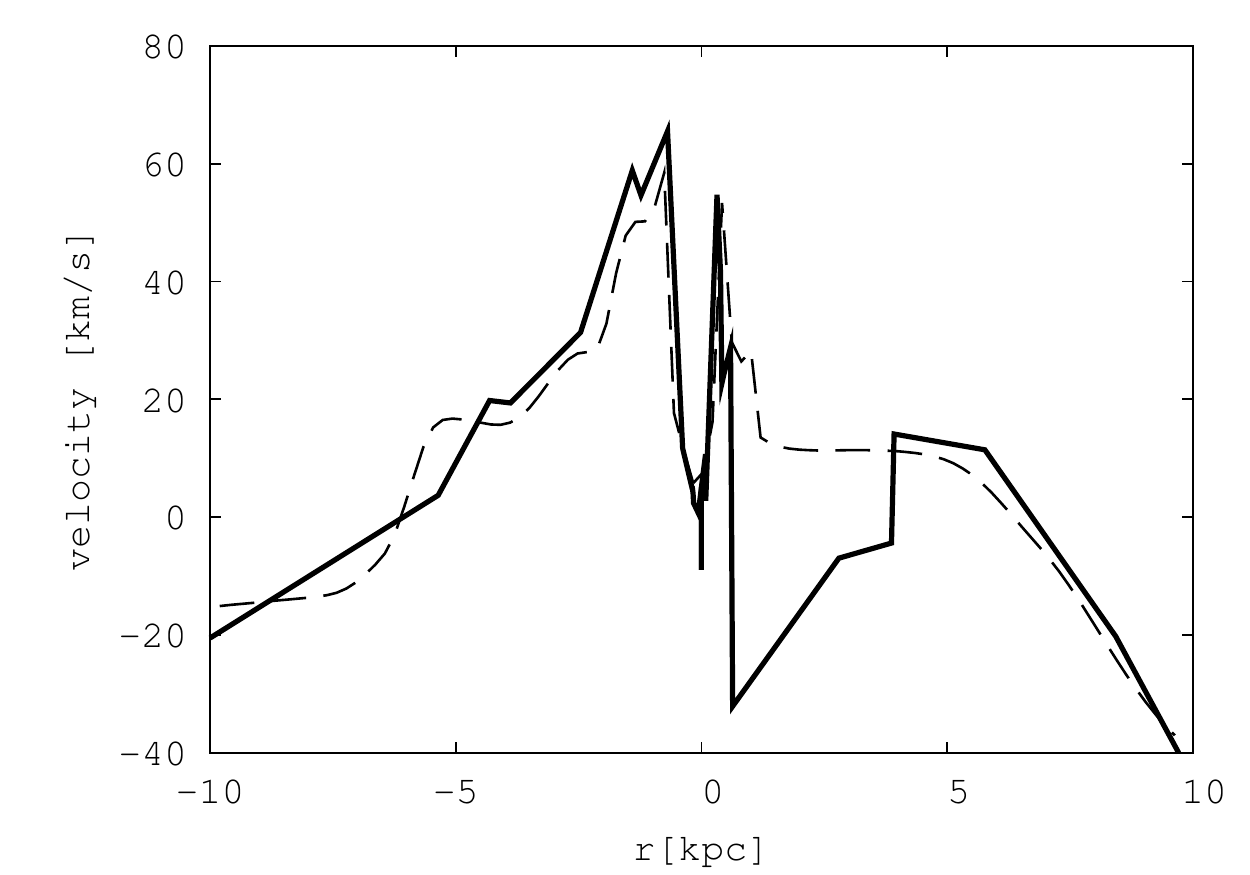}
		\end{center}
	\end{minipage}			
	\end{tabular}
	\end{center}
 	\caption{Velocity profile in a certain direction. 
		Upper panel shows the profile from the centre to the edge of the galaxy. 
		Lower panel represents the profile from a source located at $\sim 36$~kpc from the centre. 
		A solid line indicates the velocities in {\small\sf SEURAT} which are determined according to the neighbour search procedure, 
while a dashed line is the velocity in {\small\sf LICORICE}. 
	}
	\label{fig:velocity_profile}
\end{figure}

We calculate the emergent \Lya spectra for the model galaxy with {\small\sf SEURAT} and {\small\sf LICORICE}. 
In Figure \ref{fig:SEURAT_LICORICE},  the spectra calculated with the two methods are compared. 
In the upper panel, we set all velocities to zero to evaluate the difference 
between the two methods in a static problem. 
As seen in this figure, the emergent spectra agree well with each other.
This demonstrates that there is no significant difference between {\small\sf SEURAT} and {\small\sf LICORICE} 
for a static system. 
The lower panel in Fig. \ref{fig:SEURAT_LICORICE} represents the result including the velocity structure. 
In both cases, we obtain asymmetric spectra shifted to shorter wavelength, which is typical of infalling gas. 
However, the red peak is slightly weaker in the result with {\small\sf SEURAT},
and the blue peak is stronger, compared to that with {\small\sf LICORICE}. 
Since the agreement is excellent in  a static problem, we believe this discrepancy comes from 
the difference in tracing the velocity structure between the two methods. 
To verify this, we check the velocity profiles in the two methods. 
In Fig. \ref{fig:velocity_profile}, we show the profiles measured from different points. 
The upper panel represents the velocity profile from the centre to the edge of the galaxy.
We find that the two profiles behave in a fairly similar way.
The lower panel represents the profile from a certain source located at $\sim 36$~kpc from the centre. 
In this figure, we see a significant discrepancy at $0 \lesssim r \lesssim 5$~kpc,
where the velocity is negative in  {\small\sf SEURAT}, but is positive in {\small\sf LICORICE}.
This may come from the difference in the way the velocity is computed. 
In {\small\sf SEURAT},  a local SPH particle with $v < 0$ is possibly selected when assigning the velocity along the line of sight, 
while it can be smoothed out by the interpolation on the grid in {\small\sf LICORICE}. 
Such a discrepancy is anticipated to become apparent especially in low-density regions.
Actually, we have confirmed that if we dismiss the sources in low-density regions of $r> 4.5$~kpc, then
the emergent spectra fall in a good agreement. 
Thus, we speculate that the difference of the velocity assignment in low density regions
causes the discrepancy in the spectra.

\subsubsection{Validity of the ray-tracing scheme in {\small\sf SEURAT}}
As described in $\S \ref{sec:RT}$, {\small\sf SEURAT} directly uses the density at $i$-th SPH-particle, 
following the method adopted in RSPH \citep{RSPH}. 
However, the densities can be evaluated at grid-points along a ray 
by using the kernels of SPH-particles neighbouring the grid points, 
as originally implemented by \citet[][]{Kessel&Burkert00} (hereafter KB scheme). 
To see the difference between the two methods for the density assignment, 
we apply the two methods for the model galaxy and compare the optical depth along a ray. 
To clarify the difference of the density assignment, we set field velocities to be zero, 
and neglect the scattering process in this test. 
We calculate the optical depth at the line center frequency from the galactic center to the virial radius along the $X$-axis. 
\begin{figure}
	\begin{center}			
	 	\includegraphics[width=9cm,clip]{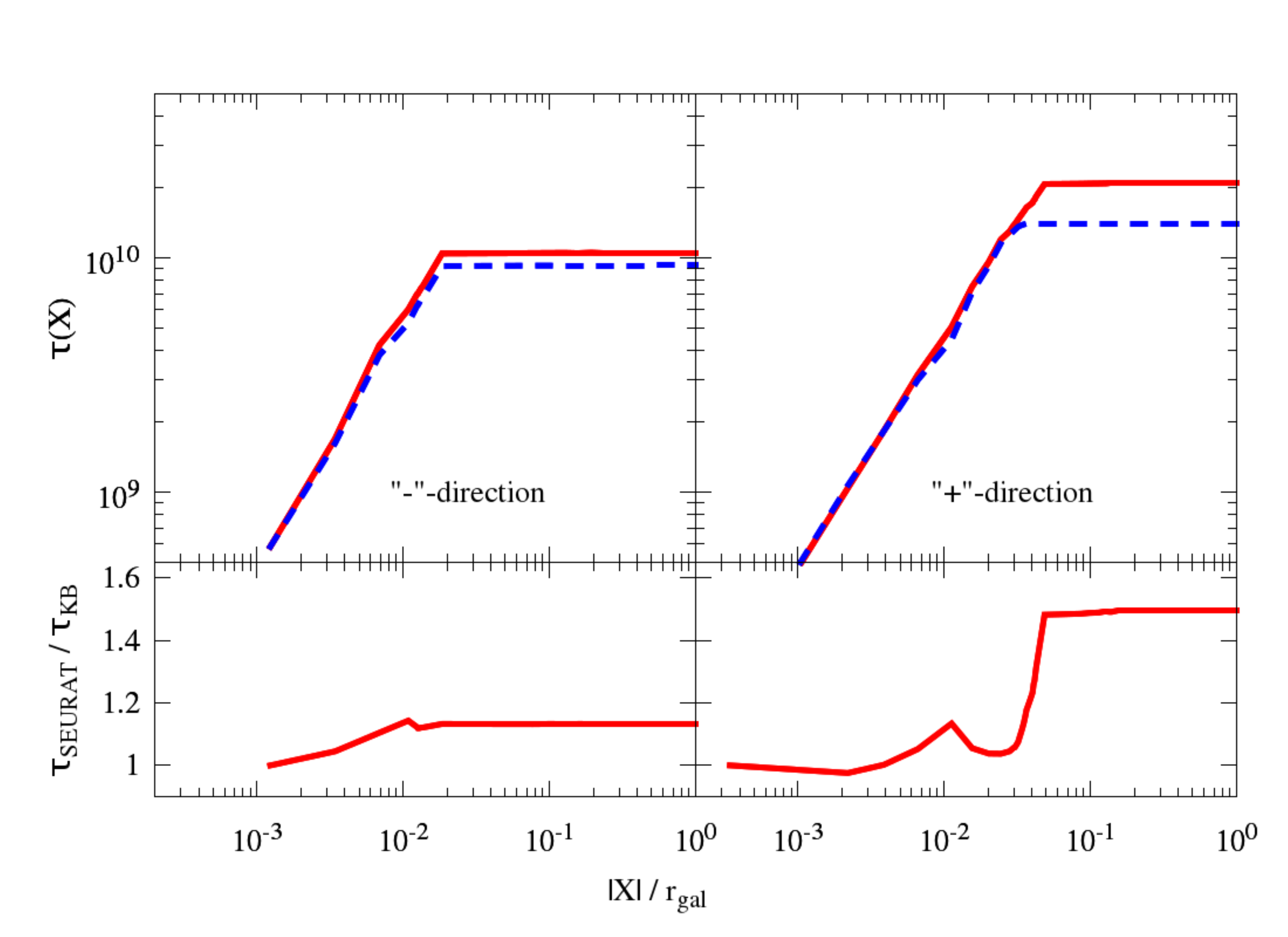}
	\end{center}
	\vspace{-12pt}
	\caption{Cumulative optical depth at the position $X$ (upper panels) and 
the ratio between {\small\sf SEURAT} and the KB scheme (lower panels). 
		The horizontal axis is the position $X$ normalized by $r_{\rm gal}$, 
where $r_{\rm gal}$ is the radius of the galaxy (twice the size of the virial radius of the halo). 
		In the upper panels, a solid line represents the result with {\small\sf SEURAT}, 
while a dashed line is that with the KB scheme. 
		The ray-tracing is performed along with the $X$-axis. 
		Left panels represent the result in the negative direction,
 while right panels display that in the positive direction. 
	}
	\label{fig:cumulative_tau}	
\end{figure}
Fig. \ref{fig:cumulative_tau} shows the resultant cumulative optical depth $\tau(X)$ as a function of position along the ray. 
Due to the high neutral hydrogen density, the optical depth steeply rises up to $\tau \sim 10^{10}$ 
in the compact inner regions of the system ($|X| / r_{\rm gal} \lesssim 10^{-2}$). 
It is notable that the two optical depths are in a good agreement in the inner regions. 
However, the agreement of $\tau(X)$ breaks down at $|X| / r_{\rm gal} \gtrsim 10^{-2}$. 
Actually, the discrepancy appears at a position where the density gradient is large. 
In {\small\sf SEURAT}, the neighbour list of particles in sparse regions does not necessarily consist of particles
in sparse regions. In other words, 
some particles can have neighbour particles both in the dense and sparse regions 
(see the particle $n$ in Fig. \ref{fig:neighbour}). 
On the other hand, it is possible that particles only in the sparse regions contribute to 
the density assignment in the KB scheme (the particle $n$ and $n'$ in Fig. \ref{fig:neighbour}). 
Thus, we see in the Fig. \ref{fig:cumulative_tau} that the cumulative optical depth calculated with {\small\sf SEURAT} 
tends to be larger than that with the KB scheme. 
However, including the information of dense particles would not be always a shortcoming. 
Increasing the number of photon packets, the possibility that a photon packet choses 
the trajectory passing through a dense region (thick dashed line in Fig. \ref{fig:neighbour})
is enhanced. 
The ray-tracing scheme in {\small\sf SEURAT} may effectively chose the trajectories
through dense regions even for the relatively small number of photon packets.
Also, it is worth mentioning that although the KB scheme seems to fit the concept of SPH, 
it would not be always accurate, since a sufficient number 
of the superposing particle is not guaranteed in the sparse regions. 
Anyway, the difference is between $1 \times 10^{10}$ and $1.5 \times 10^{10}$.
A few 10\% difference in such a large optical depth would not lead to a significant dissimilarity in 
the line profile.

\section{Conclusions}
We have developed a new numerical code, {\small\sf SEURAT}, to solve the \Lya radiative transfer adaptively based on the SPH particle distribution, using a Monte Carlo method.  
Although the previous Monte Carlo codes have been developed as mesh-based schemes, 
{\small\sf SEURAT} is a meshfree scheme, which directly uses the SPH particles themselves for the integration of optical depth. 
Hence, {\small\sf SEURAT} can solve the \Lya radiative transfer without reducing the resolution of the SPH simulations.  
We have performed the standard test calculations, which include the emergent spectra from a static homogeneous spherical cloud, the emergent spectra from an expanding homogeneous spherical cloud, and the escape fraction from a dusty slab. 
We have shown that the results reproduce the analytic solutions or the results obtained by previous studies. 
To solve \Lya radiative transfer in highly inhomogeneous media, special attention has been paid for the neighbour list construction. 
The neighbour list of the neighbour particle allows us to properly treat local optically thick regions. 
Furthermore, a scatter list is also required to perform the ray-tracing in systems having large density gradients. 
We have demonstrated that {\small\sf SEURAT} reliably searches the neighbour particles, 
and correctly performs the ray-tracing in significantly clumpy media.
As a result, we have confirmed that {\small\sf SEURAT} is successfully designed to manage extremely optically thick and highly inhomogeneous media. 
Finally, we have compared the {\small\sf SEURAT} to the mesh-based code {\small\sf LICORICE}. 
We have applied the codes to a model galaxy and calculated the emergent spectra. 
We have shown that for a static system the two methods produce very similar spectra. 
When including velocity gradients, the velocity assignment in low-density regions
may lead to some moderate differences in the emergent spectra.

So far, the SPH method has been widely employed to study galaxy formation. 
One of the potential applications of {\small\sf SEURAT} is to couple it with simulations 
of galaxy formation to model the high-$z$ LAEs. 
In modeling LAEs, both resonant scattering with the hydrogen atoms and dust scattering/absorption are essential physical processes. 
A primary advantage of {\small\sf SEURAT} is to treat such processes adaptively with the resolution of  SPH. 
We plan to carry out simulations of \Lya radiative transfer coupled with 
numerical simulations of galaxy formation. 
The results will be presented in a forthcoming paper.

\section*{Acknowledgements}
We are grateful to A. Inoue for fruitful discussions. 
The numerical simulations have been performed with COMA provided by Interdisciplinary 
Computational Science Program in Center for Computational Sciences, University of Tsukuba, 
with the K computer provided by the RIKEN Advanced Institute for Computational Science 
and with Cray XC30 at Center for Computational Astrophysics, NAOJ. 
This research was supported in part by Grant-in-Aid for Scientific Research (B) No.15H03638 (MU), 
Grant-in-Aid for Scientific Research (A) No.17H01110 (KH), 
and Grant-in-Aid for Young Scientists (A) No.17H04827 (HY) through Japan Society for the Promotion of Science.




\bibliographystyle{mnras}
\bibliography{ref-bibtex}

\begin{thebibliography}{}
\makeatletter
\relax
\def\mn@urlcharsother{\let\do\@makeother \do\$\do\&\do\#\do\^\do\_\do\%\do\~}
\def\mn@doi{\begingroup\mn@urlcharsother \@ifnextchar [ {\mn@doi@}
  {\mn@doi@[]}}
\def\mn@doi@[#1]#2{\def\@tempa{#1}\ifx\@tempa\@empty \href
  {http://dx.doi.org/#2} {doi:#2}\else \href {http://dx.doi.org/#2} {#1}\fi
  \endgroup}
\def\mn@eprint#1#2{\mn@eprint@#1:#2::\@nil}
\def\mn@eprint@arXiv#1{\href {http://arxiv.org/abs/#1} {{\tt arXiv:#1}}}
\def\mn@eprint@dblp#1{\href {http://dblp.uni-trier.de/rec/bibtex/#1.xml}
  {dblp:#1}}
\def\mn@eprint@#1:#2:#3:#4\@nil{\def\@tempa {#1}\def\@tempb {#2}\def\@tempc
  {#3}\ifx \@tempc \@empty \let \@tempc \@tempb \let \@tempb \@tempa \fi \ifx
  \@tempb \@empty \def\@tempb {arXiv}\fi \@ifundefined
  {mn@eprint@\@tempb}{\@tempb:\@tempc}{\expandafter \expandafter \csname
  mn@eprint@\@tempb\endcsname \expandafter{\@tempc}}}

\bibitem[\protect\citeauthoryear{{Ahn}, {Lee}  \& {Lee}}{{Ahn}
  et~al.}{2002}]{Ahn+02}
{Ahn} S.-H.,  {Lee} H.-W.,   {Lee} H.~M.,  2002, \mn@doi [\apj]
  {10.1086/338497}, \href {http://ads.nao.ac.jp/abs/2002ApJ...567..922A} {567,
  922}

\bibitem[\protect\citeauthoryear{{Altay}, {Croft}  \& {Pelupessy}}{{Altay}
  et~al.}{2008}]{SPHRAY}
{Altay} G.,  {Croft} R.~A.~C.,   {Pelupessy} I.,  2008, \mn@doi [\mnras]
  {10.1111/j.1365-2966.2008.13212.x}, \href
  {http://ads.nao.ac.jp/abs/2008MNRAS.386.1931A} {386, 1931}

\bibitem[\protect\citeauthoryear{{Atek}, {Kunth}, {Hayes}, {{\"O}stlin}  \&
  {Mas-Hesse}}{{Atek} et~al.}{2008}]{Atek+08}
{Atek} H.,  {Kunth} D.,  {Hayes} M.,  {{\"O}stlin} G.,   {Mas-Hesse} J.~M.,
  2008, \mn@doi [\aap] {10.1051/0004-6361:200809527}, \href
  {http://ads.nao.ac.jp/abs/2008A%26A...488..491A} {488, 491}

\bibitem[\protect\citeauthoryear{{Baek}, {Di Matteo}, {Semelin}, {Combes}  \&
  {Revaz}}{{Baek} et~al.}{2009}]{Baek+09}
{Baek} S.,  {Di Matteo} P.,  {Semelin} B.,  {Combes} F.,   {Revaz} Y.,  2009,
  \mn@doi [\aap] {10.1051/0004-6361:200810757}, \href
  {http://ads.nao.ac.jp/abs/2009A%26A...495..389B} {495, 389}

\bibitem[\protect\citeauthoryear{{Dijkstra}, {Haiman}  \& {Spaans}}{{Dijkstra}
  et~al.}{2006a}]{Dijkstra+06}
{Dijkstra} M.,  {Haiman} Z.,   {Spaans} M.,  2006a, \mn@doi [\apj]
  {10.1086/506243}, \href {http://adsabs.harvard.edu/abs/2006ApJ...649...14D}
  {649, 14}

\bibitem[\protect\citeauthoryear{{Dijkstra}, {Haiman}  \& {Spaans}}{{Dijkstra}
  et~al.}{2006b}]{Dijkstra+06b}
{Dijkstra} M.,  {Haiman} Z.,   {Spaans} M.,  2006b, \mn@doi [\apj]
  {10.1086/506244}, \href {http://ads.nao.ac.jp/abs/2006ApJ...649...37D} {649,
  37}

\bibitem[\protect\citeauthoryear{{Dijkstra}, {Lidz}  \& {Wyithe}}{{Dijkstra}
  et~al.}{2007}]{Dijkstra+07}
{Dijkstra} M.,  {Lidz} A.,   {Wyithe} J.~S.~B.,  2007, \mn@doi [\mnras]
  {10.1111/j.1365-2966.2007.11666.x}, \href
  {http://ads.nao.ac.jp/abs/2007MNRAS.377.1175D} {377, 1175}

\bibitem[\protect\citeauthoryear{{Dijkstra}, {Mesinger}  \&
  {Wyithe}}{{Dijkstra} et~al.}{2011}]{Dijkstra+11}
{Dijkstra} M.,  {Mesinger} A.,   {Wyithe} J.~S.~B.,  2011, \mn@doi [\mnras]
  {10.1111/j.1365-2966.2011.18530.x}, \href
  {http://ads.nao.ac.jp/abs/2011MNRAS.414.2139D} {414, 2139}

\bibitem[\protect\citeauthoryear{{Draine} \& {Lee}}{{Draine} \&
  {Lee}}{1984}]{Draine&Lee84}
{Draine} B.~T.,  {Lee} H.~M.,  1984, \mn@doi [\apj] {10.1086/162480}, \href
  {http://ads.nao.ac.jp/abs/1984ApJ...285...89D} {285, 89}

\bibitem[\protect\citeauthoryear{{Faucher-Gigu{\`e}re}, {Kere{\v s}},
  {Dijkstra}, {Hernquist}  \& {Zaldarriaga}}{{Faucher-Gigu{\`e}re}
  et~al.}{2010}]{Faucher-Giguere+10}
{Faucher-Gigu{\`e}re} C.-A.,  {Kere{\v s}} D.,  {Dijkstra} M.,  {Hernquist} L.,
    {Zaldarriaga} M.,  2010, \mn@doi [\apj] {10.1088/0004-637X/725/1/633},
  \href {http://ads.nao.ac.jp/abs/2010ApJ...725..633F} {725, 633}

\bibitem[\protect\citeauthoryear{{Finkelstein} et~al.,}{{Finkelstein}
  et~al.}{2013}]{Finkelstein+13}
{Finkelstein} S.~L.,  et~al., 2013, \mn@doi [\nat] {10.1038/nature12657}, \href
  {http://ads.nao.ac.jp/abs/2013Natur.502..524F} {502, 524}

\bibitem[\protect\citeauthoryear{{Genzel} et~al.,}{{Genzel}
  et~al.}{2011}]{Genzel+11}
{Genzel} R.,  et~al., 2011, \mn@doi [\apj] {10.1088/0004-637X/733/2/101}, \href
  {http://ads.nao.ac.jp/abs/2011ApJ...733..101G} {733, 101}

\bibitem[\protect\citeauthoryear{{Gronke}, {Bull}  \& {Dijkstra}}{{Gronke}
  et~al.}{2015}]{Gronke+15}
{Gronke} M.,  {Bull} P.,   {Dijkstra} M.,  2015, \mn@doi [\apj]
  {10.1088/0004-637X/812/2/123}, \href
  {http://ads.nao.ac.jp/abs/2015ApJ...812..123G} {812, 123}

\bibitem[\protect\citeauthoryear{{Harrington}}{{Harrington}}{1973}]{Harrington73}
{Harrington} J.~P.,  1973, \mn@doi [\mnras] {10.1093/mnras/162.1.43}, \href
  {http://ads.nao.ac.jp/abs/1973MNRAS.162...43H} {162, 43}

\bibitem[\protect\citeauthoryear{{Hasegawa} \& {Semelin}}{{Hasegawa} \&
  {Semelin}}{2013}]{Hasegawa&Semelin13}
{Hasegawa} K.,  {Semelin} B.,  2013, \mn@doi [\mnras] {10.1093/mnras/sts021},
  \href {http://ads.nao.ac.jp/abs/2013MNRAS.428..154H} {428, 154}

\bibitem[\protect\citeauthoryear{{Hasegawa} \& {Umemura}}{{Hasegawa} \&
  {Umemura}}{2010}]{START}
{Hasegawa} K.,  {Umemura} M.,  2010, \mn@doi [\mnras]
  {10.1111/j.1365-2966.2010.17100.x}, \href
  {http://ads.nao.ac.jp/abs/2010MNRAS.407.2632H} {407, 2632}

\bibitem[\protect\citeauthoryear{{Hernquist} \& {Katz}}{{Hernquist} \&
  {Katz}}{1989}]{Hernquist&Katz89}
{Hernquist} L.,  {Katz} N.,  1989, \mn@doi [\apjs] {10.1086/191344}, \href
  {http://ads.nao.ac.jp/abs/1989ApJS...70..419H} {70, 419}

\bibitem[\protect\citeauthoryear{{Hopkins}, {Kere{\v s}}, {O{\~n}orbe},
  {Faucher-Gigu{\`e}re}, {Quataert}, {Murray}  \& {Bullock}}{{Hopkins}
  et~al.}{2014}]{Hopkins+14}
{Hopkins} P.~F.,  {Kere{\v s}} D.,  {O{\~n}orbe} J.,  {Faucher-Gigu{\`e}re}
  C.-A.,  {Quataert} E.,  {Murray} N.,   {Bullock} J.~S.,  2014, \mn@doi
  [\mnras] {10.1093/mnras/stu1738}, \href
  {http://ads.nao.ac.jp/abs/2014MNRAS.445..581H} {445, 581}

\bibitem[\protect\citeauthoryear{{Hui} \& {Gnedin}}{{Hui} \&
  {Gnedin}}{1997}]{Hui&Gnedin97}
{Hui} L.,  {Gnedin} N.~Y.,  1997, \mn@doi [\mnras] {10.1093/mnras/292.1.27},
  \href {http://ads.nao.ac.jp/abs/1997MNRAS.292...27H} {292, 27}

\bibitem[\protect\citeauthoryear{{Iye} et~al.,}{{Iye} et~al.}{2006}]{Iye+06}
{Iye} M.,  et~al., 2006, \mn@doi [\nat] {10.1038/nature05104}, \href
  {http://ads.nao.ac.jp/abs/2006Natur.443..186I} {443, 186}

\bibitem[\protect\citeauthoryear{{Kashikawa} et~al.,}{{Kashikawa}
  et~al.}{2006}]{Kashikawa+06}
{Kashikawa} N.,  et~al., 2006, \mn@doi [\apj] {10.1086/504966}, \href
  {http://ads.nao.ac.jp/abs/2006ApJ...648....7K} {648, 7}

\bibitem[\protect\citeauthoryear{{Kessel-Deynet} \& {Burkert}}{{Kessel-Deynet}
  \& {Burkert}}{2000}]{Kessel&Burkert00}
{Kessel-Deynet} O.,  {Burkert} A.,  2000, \mn@doi [\mnras]
  {10.1046/j.1365-8711.2000.03451.x}, \href
  {http://ads.nao.ac.jp/abs/2000MNRAS.315..713K} {315, 713}

\bibitem[\protect\citeauthoryear{{Konno} et~al.,}{{Konno}
  et~al.}{2014}]{Konno+14}
{Konno} A.,  et~al., 2014, \mn@doi [\apj] {10.1088/0004-637X/797/1/16}, \href
  {http://ads.nao.ac.jp/abs/2014ApJ...797...16K} {797, 16}

\bibitem[\protect\citeauthoryear{{Laursen}, {Razoumov}  \&
  {Sommer-Larsen}}{{Laursen} et~al.}{2009}]{Laursen+09}
{Laursen} P.,  {Razoumov} A.~O.,   {Sommer-Larsen} J.,  2009, \mn@doi [\apj]
  {10.1088/0004-637X/702/1/824}, \href
  {http://ads.nao.ac.jp/abs/2009ApJ...702..824L} {702, 824}

\bibitem[\protect\citeauthoryear{{Mathis}, {Rumpl}  \& {Nordsieck}}{{Mathis}
  et~al.}{1977}]{Mathis+77}
{Mathis} J.~S.,  {Rumpl} W.,   {Nordsieck} K.~H.,  1977, \mn@doi [\apj]
  {10.1086/155591}, \href {http://ads.nao.ac.jp/abs/1977ApJ...217..425M} {217,
  425}

\bibitem[\protect\citeauthoryear{{Neufeld}}{{Neufeld}}{1990}]{Neufeld90}
{Neufeld} D.~A.,  1990, \mn@doi [\apj] {10.1086/168375}, \href
  {http://ads.nao.ac.jp/abs/1990ApJ...350..216N} {350, 216}

\bibitem[\protect\citeauthoryear{{Ono} et~al.,}{{Ono} et~al.}{2012}]{Ono+12}
{Ono} Y.,  et~al., 2012, \mn@doi [\apj] {10.1088/0004-637X/744/2/83}, \href
  {http://ads.nao.ac.jp/abs/2012ApJ...744...83O} {744, 83}

\bibitem[\protect\citeauthoryear{{Osterbrock} \& {Ferland}}{{Osterbrock} \&
  {Ferland}}{2006}]{Osterbrock&Ferland06}
{Osterbrock} D.~E.,  {Ferland} G.~J.,  2006, {Astrophysics of gaseous nebulae
  and active galactic nuclei}

\bibitem[\protect\citeauthoryear{{Ouchi} et~al.,}{{Ouchi}
  et~al.}{2008}]{Ouchi+08}
{Ouchi} M.,  et~al., 2008, \mn@doi [\apjs] {10.1086/527673}, \href
  {http://ads.nao.ac.jp/abs/2008ApJS..176..301O} {176, 301}

\bibitem[\protect\citeauthoryear{{Ouchi} et~al.,}{{Ouchi}
  et~al.}{2010}]{Ouchi+10}
{Ouchi} M.,  et~al., 2010, \mn@doi [\apj] {10.1088/0004-637X/723/1/869}, \href
  {http://ads.nao.ac.jp/abs/2010ApJ...723..869O} {723, 869}

\bibitem[\protect\citeauthoryear{{Partridge} \& {Peebles}}{{Partridge} \&
  {Peebles}}{1967}]{Partridge&Peebles67}
{Partridge} R.~B.,  {Peebles} P.~J.~E.,  1967, \mn@doi [\apj] {10.1086/149079},
  \href {http://ads.nao.ac.jp/abs/1967ApJ...147..868P} {147, 868}

\bibitem[\protect\citeauthoryear{{Pawlik} \& {Schaye}}{{Pawlik} \&
  {Schaye}}{2008}]{TRAPHIC}
{Pawlik} A.~H.,  {Schaye} J.,  2008, \mn@doi [\mnras]
  {10.1111/j.1365-2966.2008.13601.x}, \href
  {http://ads.nao.ac.jp/abs/2008MNRAS.389..651P} {389, 651}

\bibitem[\protect\citeauthoryear{{Santos}}{{Santos}}{2004}]{Santos04}
{Santos} M.~R.,  2004, \mn@doi [\mnras] {10.1111/j.1365-2966.2004.07594.x},
  \href {http://ads.nao.ac.jp/abs/2004MNRAS.349.1137S} {349, 1137}

\bibitem[\protect\citeauthoryear{{Schaye} et~al.,}{{Schaye}
  et~al.}{2015}]{Schaye+15}
{Schaye} J.,  et~al., 2015, \mn@doi [\mnras] {10.1093/mnras/stu2058}, \href
  {http://ads.nao.ac.jp/abs/2015MNRAS.446..521S} {446, 521}

\bibitem[\protect\citeauthoryear{{Semelin}, {Combes}  \& {Baek}}{{Semelin}
  et~al.}{2007}]{LICORICE}
{Semelin} B.,  {Combes} F.,   {Baek} S.,  2007, \mn@doi [\aap]
  {10.1051/0004-6361:20077965}, \href
  {http://ads.nao.ac.jp/abs/2007A%26A...474..365S} {474, 365}

\bibitem[\protect\citeauthoryear{{Shibuya}, {Kashikawa}, {Ota}, {Iye}, {Ouchi},
  {Furusawa}, {Shimasaku}  \& {Hattori}}{{Shibuya} et~al.}{2012}]{Shibuya+12}
{Shibuya} T.,  {Kashikawa} N.,  {Ota} K.,  {Iye} M.,  {Ouchi} M.,  {Furusawa}
  H.,  {Shimasaku} K.,   {Hattori} T.,  2012, \mn@doi [\apj]
  {10.1088/0004-637X/752/2/114}, \href
  {http://ads.nao.ac.jp/abs/2012ApJ...752..114S} {752, 114}

\bibitem[\protect\citeauthoryear{{Smith}, {Safranek-Shrader}, {Bromm}  \&
  {Milosavljevi{\'c}}}{{Smith} et~al.}{2015}]{Smith+14}
{Smith} A.,  {Safranek-Shrader} C.,  {Bromm} V.,   {Milosavljevi{\'c}} M.,
  2015, \mn@doi [\mnras] {10.1093/mnras/stv565}, \href
  {http://adsabs.harvard.edu/abs/2015MNRAS.449.4336S} {449, 4336}

\bibitem[\protect\citeauthoryear{{Springel}}{{Springel}}{2010}]{Springel10}
{Springel} V.,  2010, \mn@doi [\araa] {10.1146/annurev-astro-081309-130914},
  \href {http://adsabs.harvard.edu/abs/2010ARA%26A..48..391S} {48, 391}

\bibitem[\protect\citeauthoryear{{Susa}}{{Susa}}{2006}]{RSPH}
{Susa} H.,  2006, \mn@doi [\pasj] {10.1093/pasj/58.2.445}, \href
  {http://ads.nao.ac.jp/abs/2006PASJ...58..445S} {58, 445}

\bibitem[\protect\citeauthoryear{{Tasitsiomi}}{{Tasitsiomi}}{2006}]{Tasitsiomi06}
{Tasitsiomi} A.,  2006, \mn@doi [\apj] {10.1086/504460}, \href
  {http://adsabs.harvard.edu/abs/2006ApJ...645..792T} {645, 792}

\bibitem[\protect\citeauthoryear{{Vanzella} et~al.,}{{Vanzella}
  et~al.}{2011}]{Vanzella+11}
{Vanzella} E.,  et~al., 2011, \mn@doi [\apjl] {10.1088/2041-8205/730/2/L35},
  \href {http://ads.nao.ac.jp/abs/2011ApJ...730L..35V} {730, L35}

\bibitem[\protect\citeauthoryear{{Verhamme}, {Schaerer}  \&
  {Maselli}}{{Verhamme} et~al.}{2006}]{Verhamme+06}
{Verhamme} A.,  {Schaerer} D.,   {Maselli} A.,  2006, \mn@doi [\aap]
  {10.1051/0004-6361:20065554}, \href
  {http://adsabs.harvard.edu/abs/2006A%26A...460..397V} {460, 397}

\bibitem[\protect\citeauthoryear{{Verhamme}, {Schaerer}, {Atek}  \&
  {Tapken}}{{Verhamme} et~al.}{2008}]{Verhamme+08}
{Verhamme} A.,  {Schaerer} D.,  {Atek} H.,   {Tapken} C.,  2008, \mn@doi [\aap]
  {10.1051/0004-6361:200809648}, \href
  {http://ads.nao.ac.jp/abs/2008A%26A...491...89V} {491, 89}

\bibitem[\protect\citeauthoryear{{Vogelsberger} et~al.,}{{Vogelsberger}
  et~al.}{2014}]{Vogelsberger+14}
{Vogelsberger} M.,  et~al., 2014, \mn@doi [\nat] {10.1038/nature13316}, \href
  {http://ads.nao.ac.jp/abs/2014Natur.509..177V} {509, 177}

\bibitem[\protect\citeauthoryear{{Wise}, {Turk}, {Norman}  \& {Abel}}{{Wise}
  et~al.}{2012}]{Wise+12}
{Wise} J.~H.,  {Turk} M.~J.,  {Norman} M.~L.,   {Abel} T.,  2012, \mn@doi
  [\apj] {10.1088/0004-637X/745/1/50}, \href
  {http://ads.nao.ac.jp/abs/2012ApJ...745...50W} {745, 50}

\bibitem[\protect\citeauthoryear{{Yajima} \& {Li}}{{Yajima} \&
  {Li}}{2014}]{Yajima&Li14}
{Yajima} H.,  {Li} Y.,  2014, \mn@doi [\mnras] {10.1093/mnras/stu1982}, \href
  {http://ads.nao.ac.jp/abs/2014MNRAS.445.3674Y} {445, 3674}

\bibitem[\protect\citeauthoryear{{Yajima}, {Umemura}, {Mori}  \&
  {Nakamoto}}{{Yajima} et~al.}{2009}]{Yajima+09}
{Yajima} H.,  {Umemura} M.,  {Mori} M.,   {Nakamoto} T.,  2009, \mn@doi
  [\mnras] {10.1111/j.1365-2966.2009.15195.x}, \href
  {http://ads.nao.ac.jp/abs/2009MNRAS.398..715Y} {398, 715}

\bibitem[\protect\citeauthoryear{{Yajima}, {Li}, {Zhu}  \& {Abel}}{{Yajima}
  et~al.}{2012a}]{ART2}
{Yajima} H.,  {Li} Y.,  {Zhu} Q.,   {Abel} T.,  2012a, \mn@doi [\mnras]
  {10.1111/j.1365-2966.2012.21228.x}, \href
  {http://ads.nao.ac.jp/abs/2012MNRAS.424..884Y} {424, 884}

\bibitem[\protect\citeauthoryear{{Yajima}, {Li}, {Zhu}, {Abel}, {Gronwall}  \&
  {Ciardullo}}{{Yajima} et~al.}{2012b}]{Yajima+12}
{Yajima} H.,  {Li} Y.,  {Zhu} Q.,  {Abel} T.,  {Gronwall} C.,   {Ciardullo} R.,
   2012b, \mn@doi [\apj] {10.1088/0004-637X/754/2/118}, \href
  {http://ads.nao.ac.jp/abs/2012ApJ...754..118Y} {754, 118}

\bibitem[\protect\citeauthoryear{{Yajima}, {Li}, {Zhu}, {Abel}, {Gronwall}  \&
  {Ciardullo}}{{Yajima} et~al.}{2014}]{Yajima+14}
{Yajima} H.,  {Li} Y.,  {Zhu} Q.,  {Abel} T.,  {Gronwall} C.,   {Ciardullo} R.,
   2014, \mn@doi [\mnras] {10.1093/mnras/stu299}, \href
  {http://adsabs.harvard.edu/abs/2014MNRAS.440..776Y} {440, 776}

\bibitem[\protect\citeauthoryear{{Yajima}, {Li}, {Zhu}  \& {Abel}}{{Yajima}
  et~al.}{2015}]{Yajima+15}
{Yajima} H.,  {Li} Y.,  {Zhu} Q.,   {Abel} T.,  2015, \mn@doi [\apj]
  {10.1088/0004-637X/801/1/52}, \href
  {http://ads.nao.ac.jp/abs/2015ApJ...801...52Y} {801, 52}

\bibitem[\protect\citeauthoryear{{Yajima}, {Ricotti}, {Park}  \&
  {Sugimura}}{{Yajima} et~al.}{2017}]{Yajima+17}
{Yajima} H.,  {Ricotti} M.,  {Park} K.,   {Sugimura} K.,  2017, \mn@doi [\apj]
  {10.3847/1538-4357/aa8269}, \href
  {http://ads.nao.ac.jp/abs/2017ApJ...846....3Y} {846, 3}

\bibitem[\protect\citeauthoryear{{Zheng} \& {Miralda-Escud{\'e}}}{{Zheng} \&
  {Miralda-Escud{\'e}}}{2002}]{Zheng&Miralda02}
{Zheng} Z.,  {Miralda-Escud{\'e}} J.,  2002, \mn@doi [\apj] {10.1086/342400},
  \href {http://adsabs.harvard.edu/abs/2002ApJ...578...33Z} {578, 33}

\bibitem[\protect\citeauthoryear{{Zitrin} et~al.,}{{Zitrin}
  et~al.}{2015}]{Zitrin+15}
{Zitrin} A.,  et~al., 2015, \mn@doi [\apjl] {10.1088/2041-8205/810/1/L12},
  \href {http://ads.nao.ac.jp/abs/2015ApJ...810L..12Z} {810, L12}

\makeatother
\end{thebibliography}



%
%


\bsp	
\label{lastpage}
\end{document}